\documentclass[aps, prd, twocolumn, groupedaddress,showpacs]{revtex4-1}

\usepackage[english]{babel}
\usepackage[T1]{fontenc}
\usepackage[utf8]{inputenc}

\usepackage{amsmath}
\usepackage{graphicx}
\usepackage{amssymb}
\usepackage{multirow}
\usepackage{subfigure}
\usepackage{xspace}
\usepackage[dvipsnames]{xcolor}

\begin{document}

\title{Improved limits on Lorentz invariance violation from astrophysical gamma-ray sources}

\author{Rodrigo Guedes Lang}
\email{rodrigo.lang@usp.br}
\affiliation{Instituto de F\'isica de S\~ao Carlos, Universidade de S\~ao Paulo, Av. Trabalhador S\~ao-carlense 400, S\~ao Carlos, SP, Brasil.}

\author{Humberto Mart\'inez-Huerta}
\email[]{humbertomh@ifsc.usp.br} 
\affiliation{Instituto de F\'isica de S\~ao Carlos, Universidade de S\~ao Paulo, Av. Trabalhador S\~ao-carlense 400, S\~ao Carlos, SP, Brasil.}

\author{Vitor de Souza}
\email[]{vitor@ifsc.usp.br}
\affiliation{Instituto de F\'isica de S\~ao Carlos, Universidade de S\~ao Paulo, Av. Trabalhador S\~ao-carlense 400, S\~ao Carlos, SP, Brasil.}

\date{February 27, 2019}

\begin{abstract}
Lorentz invariance (LI) has a central role in science and its violation (LIV) at some high-energy scale has been related to possible solutions for several of the most intriguing puzzles in nature such as dark matter, dark energy, cosmic rays generation in extreme astrophysical objects and quantum gravity. We report on a search for LIV signal based on the propagation of gamma rays from astrophysical sources to Earth. An innovative data analysis is presented which allowed us to extract unprecedented information from the most updated data set composed of 111 energy spectra of 38 different sources measured by current gamma-ray observatories. No LIV signal was found, and we show that the data are best described by LI assumption. We derived limits for the LIV energy scale at least 3 times better than the ones currently available in the literature for subluminal signatures of LIV in high-energy gamma rays.
\end{abstract}

\pacs{11.30.Cp, 98.70.Rz, 95.30.Cq}

\maketitle

\section{Introduction}
\label{sec:introduction}

Lorentz invariance (LI) is one of the pillars of fundamental physics and its violation (LIV) has been proposed by several quantum gravity and effective field theories~\cite{NAMBU, Kostelecky:1988zi, Colladay:1998fq, QG4,QG5, QG1,  ALFARO, Pot, Audren:2013dwa, Bluhm,Calcagni:2016zqv,  Bettoni:2017lxf}.
Astroparticles have proven to be a sensitive probe for LIV and its signatures in the photon sector have been searched through arrival time delay, photon splitting, spontaneous emission, shift in the pair production energy threshold and many others effects~\cite{bib:liv:tests:astropart, Coleman:1997xq, AmelinoEllis:1998, Coleman:1998ti, Stecker:2001vb, Stecker:2003pw, Jacobson:2002, Stecker:2004, Ellis2006402, Gunter:2007, Gunter:2008, ALBERT2008253,    Stecker:2009,   XU201672,1674-1137-40-4-045102, ELLIS201350,   Farbairn:2014,  Tavecchio:2015, Biteau:2015, Martinez-Huerta:2016azo, Rubtosov:2017, Martinez-Huerta:2017gna, Lang:2017wpe,  Cologna:2016cws, Mrk501_HESS_flare,  Pfeifer:2018pty,  Ellis:2018lca,Abdalla:2018sxi}.
In particular, the strongest limits for subluminal signatures of LIV based on the propagation of high-energy gamma rays have been imposed using the energy spectra of TeV gamma rays sources~\cite{Mrk501_HESS_flare,Cologna:2016cws,Biteau:2015} and the time delay of TeV gamma-rays emitted by gamma-ray bursts (GRBs)~\cite{Vasileiou:2013vra}.

The framework summarized in the next section shows how the interaction of gamma rays with background photons on the way from the source to Earth modulates the intrinsic energy spectrum emitted by the source. The modulation of the spectrum is considerably different if the interactions in the propagation are taken to be LI or LIV. Previous works have shown how to extract the effect of the propagation from the measured energy spectrum allowing us to identify the assumption (LI or LIV) which best describes the data~\cite{Mrk501_HESS_flare,Cologna:2016cws,Biteau:2015}. These analyses have been limited mainly by (a) poor knowledge of the extra-Galactic background light (EBL), (b) large uncertainties in the intrinsic energy spectra functional form, (c) scarce data and (d) not fully optimized analysis procedures.

In this work, a new analysis method is proposed to help overcoming these limitations and to contribute in improving the power to search for LIV signatures in TeV gamma-ray energy spectra. Moreover, the most updated data set available in the literature was analyzed: 111 energy spectra from 38 different sources. Two selection steps are implemented in this analysis. First, a selection procedure is developed to choose the relevant measured spectra. We show that only 18 spectra from 6 sources out of the 111 spectra from 38 sources have power to constrain LIV beyond the current limits. This selection procedure developed here can be used in any future analysis to evaluate which new measured spectrum is relevant to impose LIV limits. Second, the analysis method developed in Sec. III considers carefully each measured point of each spectrum, rejecting any data that could bias the result towards a faked positive LIV signal. The use of the most complete data set combined with an innovative analysis procedure resulted in the best LIV limits derived so far using this framework as shown in  Sec. IV. The limitations of the method developed here are tested in Appendix~\ref{sec:systematics} in which we shown that the results are robust
under (a) poor knowledge of the EBL, (b) large uncertainties in the intrinsic energy spectra functional form, (c) energy resolution, (d) selection of spectra and (e) energy bins selection used in the calculation of the intrinsic energy spectra. 

\section{LIV in the gamma-ray astrophysics framework}
\label{sec:framework}

Subluminal LIV in the photon sector can be described as a polynomial correction of the dispersion relation:

\begin{equation}
\label{eq:LIV}
E^2_{\gamma} - p^2_{\gamma} = -  \frac{E_\gamma^{(n+2)}}{\left(E_{\mathrm{LIV}}^{(n)}\right)^n},
\end{equation}
where $E_\gamma$ is the energy and $p_\gamma$ the momentum of the gamma ray. Natural units are used in this work (c = 1). $E_{\mathrm{LIV}}^{(n)}$ is the LIV energy scale for each correction order $n$.  $E_{\mathrm{LIV}}^{(n)}$ is the parameter to be constrained in this analysis because it modulates the effect and is used to derive the energy beyond which the energy dispersion relation departures from LI. Only the two leading orders $n=$ 1, 2 are considered hereafter. Best current limits at 2$\sigma$ confidence level for subluminal signatures of LIV in high-energy gamma rays are $E_{\mathrm{LIV}}^{(1)} = 9.27\times 10^{28}$~eV~\cite{Vasileiou:2013vra} and $E_{\mathrm{LIV}}^{(2)} =  8.7\times10^{20}$~eV~\cite{Cologna:2016cws}.

On their way to Earth, TeV gamma rays interact with the EBL photons creating pairs:\newline $\gamma + \gamma_{EBL} \rightarrow e^{+} + e^{-}$~\cite{DeAngelis:2013jna}. In the next section we use the EBL model of Franceschini~\cite{Franceschini} as implemented in reference~\cite{ebltable} and the influence of other models~\cite{Gilmore,Dominguez} is tested in Appendix~\ref{sec:Ab_EBL}. Successive interactions attenuate the emitted gamma-ray flux as described by
\begin{equation}
  a (E,z) = e^{-\tau} (E,z) = \frac{J_{meas} (E) }{J_{int} (E,z)},
  \label{eq:attenuation}
\end{equation}
where $J_{meas}$ is the measured spectrum at Earth and $J_{int}$ is the intrinsic spectrum emitted by the source. $a(E,z)$ is called attenuation and $\tau$ is the optical depth.

\begin{figure}
  \centering
  \includegraphics[width=0.48\textwidth]{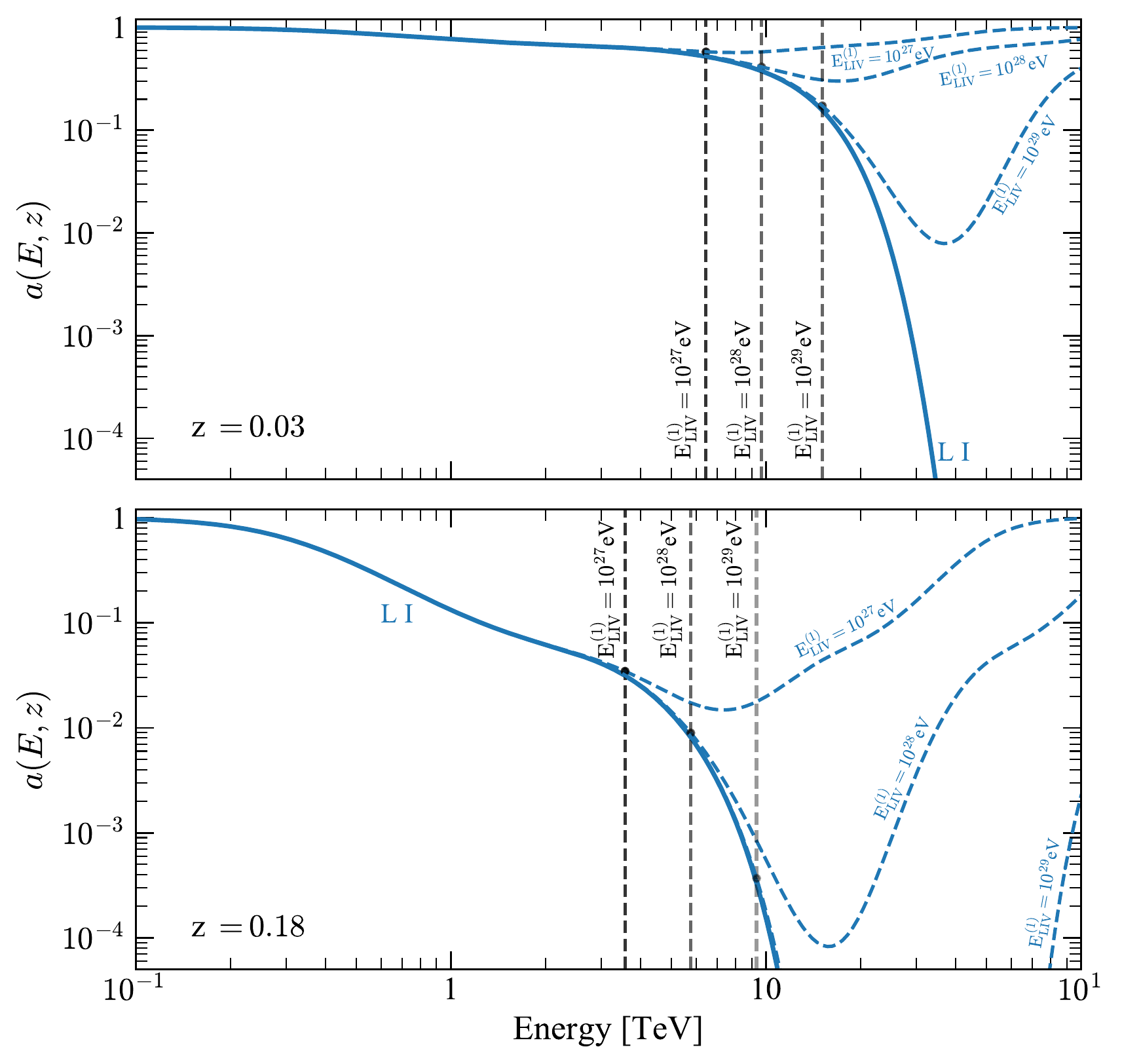}
  \caption{Examples of attenuation as a function of gamma-ray energy. Upper and bottom panels correspond to a source at z = 0.03 and 0.18, respectively. Four cases are shown for LI and LIV with $E_{\mathrm{LIV}}^{(1)} = 10^{27}, \; 10^{28}$ and $10^{29}$ eV. Dashed vertical lines shows the energy in which the LIV attenuation is 10\% higher than the LI attenuation.}
  \label{fig:attenuation}
\end{figure}

If LIV is considered, the pair-production energy threshold increases and the gamma rays have less probability to interact with the EBL photons. As a consequence, the optical depth decreases and the gamma ray propagates farther in the Universe~\cite{DeAngelis:2013jna,Lang:2017wpe}. Figure~\ref{fig:attenuation} shows an example of this effect in the attenuation of gamma rays from two sources at z = 0.03 and 0.18 as a function of the energy. Four cases were calculated assuming LI and LIV with $E_{\mathrm{LIV}}^{(1)} = 10^{27}, \; 10^{28}$ and $10^{29}$ eV. The interaction suppresses the flux at the highest energies. When LI is considered, there is a steep and definitive drop in the attenuation curve. However, when LIV is considered, the interaction becomes less probable for the highest energetic gamma rays which can propagate further causing a recovery of the flux. The intensity of the effect depends on the energy of the gamma ray, on the LIV energy scale and on the distance of the source. In the next section, a method is developed to deal with these dependencies and extract the LIV energy scale ($E_{\mathrm{LIV}}^{(n)}$) which best describes the data.

\section{Analysis method and data selection}
\label{sec:analysis}

The use of multiple sources at different distances can be combined by proper analysis methods to improve the search for a LIV signal. Each measured energy spectrum contributes with a given strength to the analysis efficiency. Selecting only the relevant data helps to increase the statistics power without adding systematic effects. We considered here every energy spectrum measured by each observatory as independent measurements.

Figure~\ref{fig:attenuation} shows that there is an energy window of interest in between the abrupt fall of the LI attenuation and the recovery in the LIV curve ranging from a few up to hundreds of TeV depending on the source distance. In this energy range it is easier to differentiate the LI from the LIV assumption. Gamma-ray observatories have continuously coverage from hundreds of GeV to few TeV and the upper energy threshold is given by collection area. Therefore LIV studies are usually limited by the maximum energy which can be measured by the experiments.

In summary, only two quantities determine the contribution of a measured spectrum in searching for a LIV signal: the distance of the source and the maximum energy measured in that spectrum,  ($E_{max}$). The distance of the source controls the amount of modulation in the spectrum and $E_{max}$ sets how much data is available in the energy window of interest. Based on this discussion, we propose to select measurements in which the attenuation ratio between LIV and LI assumptions at $E_{max}$ differs by at least 10\%: $a_{LIV}/a_{LI} > 1.1$. These points are illustrated by the horizontal dashed lines in Fig.~\ref{fig:attenuation}.

\begin{figure}
  \centering
  \includegraphics[width=0.48\textwidth]{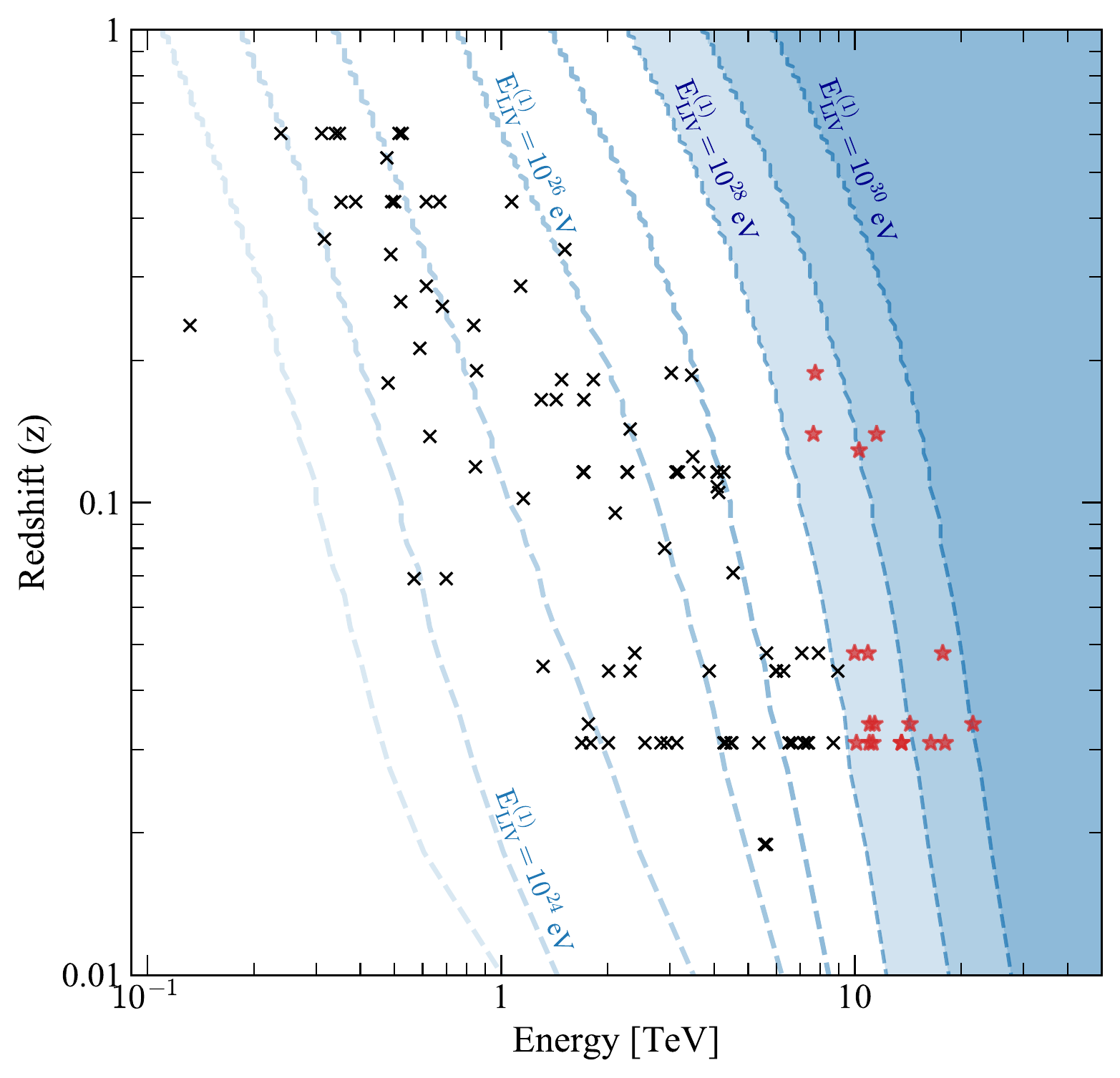}
  \caption{Distance as a function of energy for which $a_{LIV}/a_{LI} > 1.1$ are shown as dashed lines for several $E_{\mathrm{LIV}}^{(1)}$ values. Black crosses and the red stars show the distance and $E_{max}$ of all 111 measured energy spectra used in this work as taken from the TeVCat catalog~\cite{tevcat}. Red stars show the spectra with power to set a LIV energy scale limit more stringent than the current available one. The selected spectra are shown in Appendix~\ref{sec:Aa_src}.}
  \label{fig:liv:phase}
\end{figure}

Figure~\ref{fig:liv:phase} summarizes the spectrum selection procedure. Dashed curves show the distances (z) as a function of energy for which $a_{LIV}/a_{LI} > 1.1$. Each curve was calculated for a different value of $E_{\mathrm{LIV}}^{(1)}$ as shown. The black crosses and the red stars show the distance of the source and $E_{max}$ of all 111 measured energy spectra used in this work. The measurements were taken from the TeVCat catalog~\cite{tevcat} and confirmed in the original publications. An energy spectrum is useful to set a limit of $E_{\mathrm{LIV}}^{(1)}$ if its ($E_{max}$,~z) point is on the right side of the corresponding $E_{\mathrm{LIV}}^{(1)}$ line. Given that we aim at setting limits on $E_{\mathrm{LIV}}^{(1)}$ more stringent than the ones already available in the literature ($E_{\mathrm{LIV}}^{(1)} \sim 10^{28}$~eV), we have selected only 18 spectra from 6 sources~\cite{Mrk421_HEGRA_1999,Mrk421_HESS_2004,Mrk421_VERITAS_low,Mrk421_TACTIC_2005,Mrk421_TACTIC_2009,Mrk501_TACTIC,Mrk501_ARGO_flare,Mrk501_HESS_flare,s1ES1959_Whipple,s1ES1959_HEGRA_low,H1426_HEGRA_1999,s1ES0229_HESS,s1ES0229_VERITAS,s1ES0347_VERITAS}, shown as red stars in Fig.~\ref{fig:liv:phase} for further considerations in this analysis. 
In Appendix~\ref{sec:Aa_src}, these selected spectra are discriminated in Table II, and in Appendix~\ref{sec:Ab_src} we also show the effect of including other 11 spectra from 3 sources~\cite{Mrk421_VERITAS_low,Mrk421_ARGO_flux1,s1ES2344_Whipple_2005_b,s1ES2344_VERITAS_low,s1ES2344_VERITAS_2017,s1ES1959_VERITAS,s1ES1959_VERITAS_2015} in the analysis and we prove our hypothesis that the latter are useless to improve the search for a LIV signal.

Once the relevant energy spectra are chosen, the standard analysis procedure follows four steps: (I) from the measured spectra, calculate the intrinsic spectra using the LI attenuation at the distance of each source, (II) model the intrinsic spectra with a functional form, (III) using the model of the intrinsic spectra calculate the spectra at Earth supposing several LIV energy scales, (IV) compare the calculated LIV spectra on Earth with the measured spectra on Earth and set which value of LIV energy scale best describes the data.

We consider here a more carefully analysis of the first step. If LIV is true in nature, the highest energetic gamma rays measured on Earth interacted in its way under LIV, therefore, the assumption of step (I) can be false. In other words, the measured spectrum can not be deattenuated under LI assumption to calculate the intrinsic spectrum if finding a LIV signal is the target of the analysis. Apparently, this point has been neglected in previous studies of this kind and its consequence is to artificially produce a LIV signal or artificially improve the LIV energy scale limits.

We have taken care of that by using only points in each energy spectrum which could not be differentiated between a LI and LIV propagation as explained below. We define a fiducial LI region in each measured energy spectrum as the energy range in which the measured flux cannot distinguish between LI and LIV propagation. The fiducial LI region of each spectrum is constituted by the set of energy bins that fulfills the following condition:
\begin{equation}
  \frac{a_{\mathrm{LIV}}}{a_{\mathrm{LI}}} \le \frac{J_{meas} (E) + \rho \; \sigma \left(J_{meas} (E) \right)}{J_{meas} (E)},
  \label{eq:li:region}
\end{equation}
where $a_{\mathrm{LIV}}$ and $a_{\mathrm{LI}}$ are the LIV and LI attenuation, respectively. $J_{meas}$ and $\sigma \left(J_{meas}\right)$ are the measured flux and its statistical uncertainty, respectively. $\rho$ is an input parameter of the analysis taken as $\rho = 1$. In Appendix~\ref{sec:Ab_bin} we test larger values of $\rho$ and show that the results presented here are robust under reasonable choices of $\rho$.

According to the condition in equation~\ref{eq:li:region}, bins in which the difference in $a_{\mathrm{LIV}}$ and $a_{LI}$ are larger than the statistical uncertainty of the measured flux are discarded in the reconstruction of the intrinsic spectrum. Only points satisfying the condition are used to calculate the intrinsic spectrum. These points can be safely deattenuated using $a_{LI}$ (analysis step I) to calculate the intrinsic flux at the source avoiding the introduction of spurious LIV signal.

We modeled the intrinsic spectrum at the source by a simple power law and by a power law with an exponential cutoff (analysis step II). As shown in Appendix~\ref{sec:Ab_int}, the data are better described when a power law with an exponential cutoff:
\begin{equation}
  J_{int} (E) = A \left(\frac{E}{E_0}\right)^{-\Gamma} e^{-E/E_{cut}},
  \label{eq:PLEC}
\end{equation}
where the normalization $A$, the spectral index $\Gamma$ and the energy cutoff $E_{cut}$ are free parameters. $E_0$ is a reference energy taken to be $E_0 = 1$~TeV. The best fitted parameters ($A,\; \Gamma$ and $E_{cut}$) and its one sigma statistical uncertainties are considered in the next step of the analysis.

The fitted intrinsic spectra are propagated back to Earth under the assumption of LIV (analysis step III). The calculated energy spectra on Earth is defined as: $J_{cal} =  a_{LIV} \times J_{int}$ for several LIV energy scales. We varied $E_{\mathrm{LIV}}^{(1)}$ from $4 \times 10^{27}$~eV to $10^{30}$~eV in log steps of 0.0041 and $E_{\mathrm{LIV}}^{(2)}$ from $2 \times 10^{20}$~eV to $10^{22}$~eV in log steps of 0.0041. Note that the LI scenario corresponds to $E_{\mathrm{LIV}}^{(n)} \rightarrow \infty$. The most important experimental feature in the measured energy spectrum is the energy resolution of the detection. We have considered an energy resolution of 10\%. Each bin in $J_{cal}$ was smeared by a Gaussian with width equal to 10\% of the bin energy using a forward-folding technique. In Appendix~\ref{sec:Ab_ener} we tested other values of the energy resolution and show that the conclusions presented here are not changed if reasonable values of the energy resolution are considered.

At this point in the analysis, for each of the 18 measured spectra we have calculated several $J_{cal}$ spectra covering (a) many $E_{\mathrm{LIV}}^{(n)}$ values and (b) many possibilities of intrinsic spectra inside the one sigma uncertainty of the best fitted values. Each one of these $J_{cal}$ spectra is finally compared to the measured $J_{meas}$ spectra using a log-likelihood statistical test (analysis step IV). In this test, all measured points in the energy spectra are used. For each $E_{\mathrm{LIV}}^{(n)}$, the log-likelihood value ($2\mathcal{L}$) of all 18 spectra are summed. It is only here that all 18 spectra contribute together to limit one value of $E_{\mathrm{LIV}}^{(n)}$. Upper limits in the measured flux have also been used in the log-likelihood calculation and they play a very important role in limiting the recover of the LIV flux.

Without loss of generality, we have chosen to analyze only the two limiting cases within the one sigma uncertainty best fitted parameters of the intrinsic spectra. We show only the bracketing solutions of the intrinsic spectra which have the lowest and highest values of $2\mathcal{L}$. We named these solutions LIV-disfavored and LIV-favored, respectively. The variation of $2\mathcal{L}$ with $E_{\mathrm{LIV}}^{(n)}$ determines the presence of a LIV signal or the LIV energy scale limits as analyzed in details in the next section.

\section{New LIV limits}
\label{sec:results}

\begin{figure}
  \centering
  \includegraphics[width=0.48\textwidth]{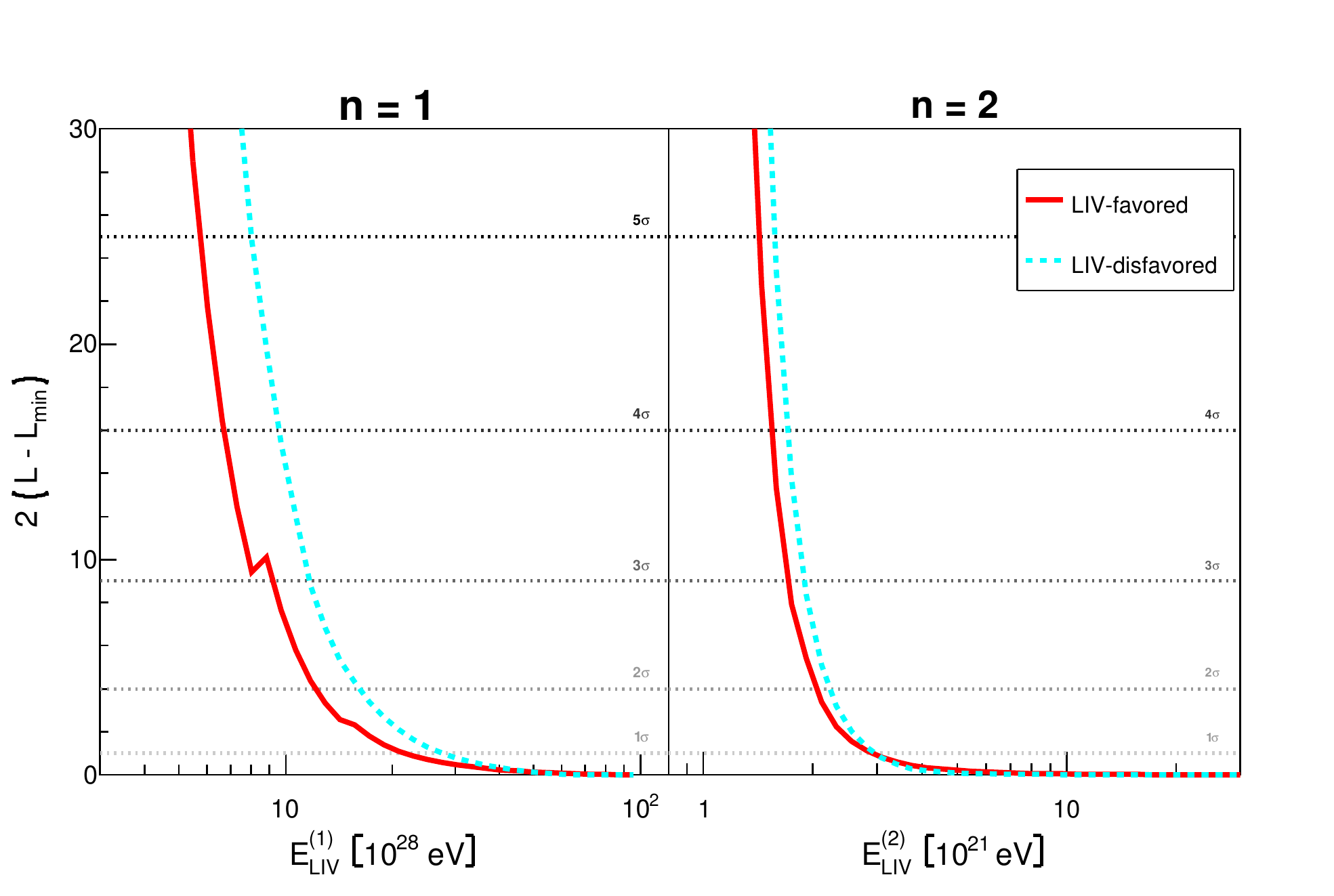}
  \caption{Log-likelihood value as a function of the LIV energy scale. Left plot is for $n=1$ and right plot for $n=2$. The red full and cyan dashed line represent the LIV-favored and LIV-disfavored cases, respectively. The horizontal dashed lines represent the 1, 2, 3, 4 and 5$\sigma$ rejection confidence levels.}
  \label{fig:result}
\end{figure}

Figure~\ref{fig:result} shows the variation of the log-likelihood value with $E_{\mathrm{LIV}}^{(n)}$ for $n=1$ and $n=2$ for LIV-disfavored and LIV-favored cases. The discontinuity in one curve in Fig.~\ref{fig:result} is a consequence of the analysis procedure explained in the last section. The discontinuity is caused by the inclusion of an extra point in the fiducial LI region when the $E_{\mathrm{LIV}}^{(n)}$ moves upwards. The minimum log-likelihood value ($\mathcal{L}_{min}$) was found when the maximum $E_{\mathrm{LIV}}^{(n)}$ is considered. The tendency of $2(\mathcal{L}-\mathcal{L}_{min})$ in Fig.~\ref{fig:result} shows the log-likelihood difference vanishing with $E_{\mathrm{LIV}}^{(n)} \rightarrow \infty$ which corresponds to the LI case. In conclusion, the data set formed by the 18 energy spectra considered here is best described by a LI model.

Thus, it is possible to impose limits on the LIV energy scale. The LIV model corresponding to a given $E_{\mathrm{LIV}}^{(n)}$ can be excluded with a confidence level (CL) given by $\sigma = \sqrt{2(\mathcal{L}-\mathcal{L}_{min})}$ as shown by the dashed horizontal lines in Fig.~\ref{fig:result}. Table~\ref{tab:limits} shows the limits imposed by this analysis for $E_{\mathrm{LIV}}^{(1)}$ and $E_{\mathrm{LIV}}^{(2)}$ with $2\sigma$, $3\sigma$ and $5\sigma$ CL. We show the limits for most conservative scenario based on the LIV-favored case.

\begin{figure}[t]
  \centering
  \includegraphics[width=0.46\textwidth]{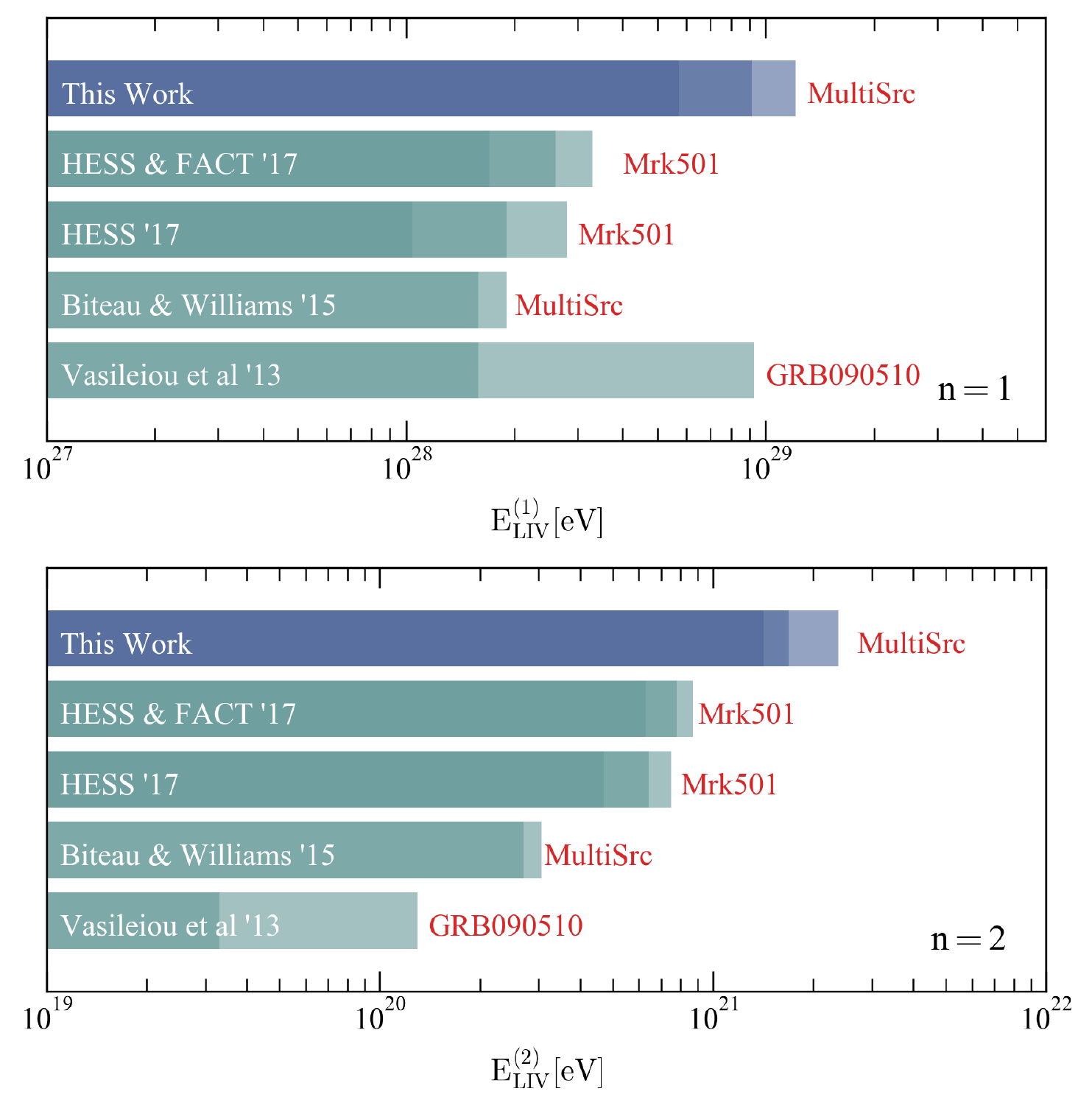}
  \caption{Comparison of the best limits imposed on the LIV energy scale. Left panel for $n=1$ and right panel for $n=2$. Shades of blue and green correspond to 2, 3 an 5 $\sigma$ CL. Only the most conservative limits of our analysis are shown corresponding to the LIV-favored case. This work and Biteau and Williams's 15 are based on multiple sources (MultiSrc), the latter of which are translated to the photon sector and to the quadratic term. The other limits are based on specific measurement of one source as appointed.}
  \label{fig:limits}
\end{figure}

Figure~\ref{fig:limits} compares the LIV energy scale limits presented in this work with the best limits in the literature: (a) the best limits from spectral analysis of a single TeV source imposed by the Markarian 501 measurements from HESS and FACT~\cite{Cologna:2016cws,Mrk501_HESS_flare}, (b) the best limits from spectral analysis of multiple TeV sources~\cite{Biteau:2015}, and (c) the best limits from time delay analysis of gamma-ray bursts (GRB) imposed by the GRB090510 measurements from MAGIC~\cite{Vasileiou:2013vra}. 
The limits imposed in this work are at least 3 times better than the ones presented in previous works.

\begin{table}
  \centering
  \begin{tabular}{c | c c c}
    \hline \hline
    & $2\sigma$ & $3\sigma$ & $5\sigma$ \\ \hline
    $E_{\mathrm{LIV}}^{(1)} \left[10^{28} \ \mathrm{eV}\right]$ & 12.08 & 9.14 & 5.73 \\
    $E_{\mathrm{LIV}}^{(2)} \left[10^{21} \ \mathrm{eV}\right]$ & 2.38 & 1.69 & 1.42 \\ \hline \hline
  \end{tabular}
  \caption{Limits on the LIV energy scale imposed by this analysis. Only the most conservative limits are shown corresponding to the LIV-favored case.}
  \label{tab:limits}
\end{table}

The comparison between the results presented here and limits obtained from the nonobservation of ultra-high-energy (UHE) photons is not straightforward. LIV limits imposed with TeV gamma rays and UHE photons are independent and complementary analysis. Some analysis using upper limits on UHE photon photon flux are more constraining than the TeV gamma-ray limits shown here, however they are strongly dependent on the astrophysical assumptions considered~\cite{Lang:2017wpe}. 

\section{Conclusions}
\label{sec:conclusions}

In this work, we propose a new analysis procedure for searching LIV signatures using multiple TeV measured energy spectra. The analysis method developed here includes (a) a procedure to select the relevant measured spectra and (b) a procedure to select which bins in each measured energy spectrum should be considered to calculate the intrinsic energy spectrum of the source. Both selections minimized the systematic bias of the analysis and allowed us to obtain a very robust result irrespective of the issues which traditionally penalized the LIV studies, such as (a) poor knowledge of the EBL, (b)  large uncertainties in the intrinsic energy spectra functional form, (c) scarce data and (d) energy resolution. The influence of these limitations and the possible biases introduced by the new criteria are evaluated in Appendix~\ref{sec:systematics}, where we show that our conclusions are valid despite these limitations and biases.

Throughout the paper we consider only subluminal LIV in the photon sector to allow the comparison with a large set of previous studies. If LIV in the electron sector is also considered (pair-production), the LIV parameter $1/{\rm E_{LIV}}$ becomes a linear combination of the LIV contributions from the different particle species \cite{Martinez-Huerta:2016azo}. In the most common scenario, photons dominate over electrons, and the derived results in this work remain the same. In the second most common scenarios, LIV is universal for photons and electrons, and a factor of $1/(1-1/2^n)$ should be considered in the final results~\cite{Martinez-Huerta:2019ehp}. The superluminal propagation of photons is not considered in the paper because its consequences would require a specific data analysis, probably different from the one used in this paper.

We applied this analysis method to the most updated gamma-ray TeV data set. We considered 111 measured energy spectra from 38 sources; only 18 measured spectra from 6 sources were shown to significantly contribute to restricting the LIV energy scale beyond the current limits. We conclude that the data set is best described by LI assumption and we impose strict limits to the LIV energy scale. Figure~\ref{fig:limits} summarizes the results. At $5 \sigma$ exclusion CL, the LIV energy scale limits imposed here are 3.3 times better than the best limits from previous TeV spectra analysis and 3.6 times better than the best limits from previous time delay analysis.


\begin{acknowledgments}

The authors acknowledge FAPESP support No. 2015/15897-1, No. 2016/24943-0 and No. 2017/03680-3. The authors also acknowledge the National Laboratory for Scientific Computing (LNCC/MCTI, Brazil) for providing HPC resources of the SDumont supercomputer, which have contributed to the research results reported within this paper (http://sdumont.lncc.br).
\end{acknowledgments}


\appendix
\section{Selected spectra}
\label{sec:Aa_src}

This section contains the list of sources used in this analysis. 

\begin{table}[h]
  \centering
  \begin{tabular}{c c c c c}
    \hline \hline
    Source & Redshift & Experiment & Spectrum & Reference \\ \hline
    \multirow{7}{*}{Markarian 421} & \multirow{7}{*}{0.031} & \multirow{2}{*}{HEGRA} & 1999-2000 & \cite{Mrk421_HEGRA_1999} \\
     &  &  & 2000-2001 & \cite{Mrk421_HEGRA_1999} \\
     &  & HESS & 2000 & \cite{Mrk421_HESS_2004} \\
     &  & \multirow{2}{*}{VERITAS} & {\tiny 2006-2008 (low)} & \cite{Mrk421_VERITAS_low} \\
     &  &  & {\tiny 2006-2008 (mid)} & \cite{Mrk421_VERITAS_low} \\
     &  & \multirow{2}{*}{TACTIC} & 2005-2006 & \cite{Mrk421_TACTIC_2005} \\
     &  &  & 2009-2010 & \cite{Mrk421_TACTIC_2009} \\ \hline
    \multirow{4}{*}{Markarian 501} & \multirow{4}{*}{0.034} & TACTIC & 2005-2006 & \cite{Mrk501_TACTIC} \\
     &  & \multirow{2}{*}{ARGO-YBJ} & 2008-2011 & \cite{Mrk501_ARGO_flare} \\
     &  &  & 2011 (flare) & \cite{Mrk501_ARGO_flare} \\
     &  & HESS & 2014 (flare) & \cite{Mrk501_HESS_flare} \\ \hline
    \multirow{3}{*}{1ES 1959+650} & \multirow{3}{*}{0.048} & Whipple & 2002 (flare) &  \cite{s1ES1959_Whipple} \\
     &  & \multirow{2}{*}{HEGRA} & 2002 (low) & \cite{s1ES1959_HEGRA_low} \\
     &  &  & 2002 (high) & \cite{s1ES1959_HEGRA_low} \\ \hline
    H 1426+428 & 0.129 & HEGRA & 1999-2000 & \cite{H1426_HEGRA_1999} \\ \hline
    \multirow{2}{*}{1ES 0229+200} & \multirow{2}{*}{0.1396} & HESS & 2005-2006 & \cite{s1ES0229_HESS} \\
     &  & VERITAS & 2010-2011 & \cite{s1ES0229_VERITAS} \\ \hline
    1ES 0347-121 & 0.188 & VERITAS & 2006 & \cite{s1ES0347_VERITAS}\\ \hline \hline
  \end{tabular}
  \caption{List of spectra 18 spectra from 6 sources out of the 111 spectra from 38 sources that fulfill the selection criteria according to the procedure proposed in the paper. Columns show the name and redshift of the source, the experiment that measured the spectra and a specification of the measured spectrum as in the original publication shown as reference in the last column.}
  \label{tab:sources}
\end{table}

\section{Evaluation of the systematic effects and of the limitations of this analysis}
\label{sec:systematics}

In this section, we evaluate the influence of systematic effects and of the limitations of this analysis in the conclusions of the work. We compare different possible choices to the benchmark model presented in Sec. IV. The main sources of systematics and limitations of the analysis are

\begin{itemize}
\item{Choices of the EBL models;}
\item{Model of the intrinsic spectrum;}
\item{Energy resolution;}
\item{Selection of spectra;}
\item{Selection of energy bins to be used in the calculation of the intrinsic energy spectra.}
\end{itemize}

The three first items are common to all LIV analysis based on the propagation of TeV gamma rays. The last two items are particular to the method proposed here. Below, we show the impact of different choices in the results.

\subsection{EBL models}
\label{sec:Ab_EBL}

\begin{figure}[bh!]
  \centering
  \includegraphics[width=0.5\textwidth]{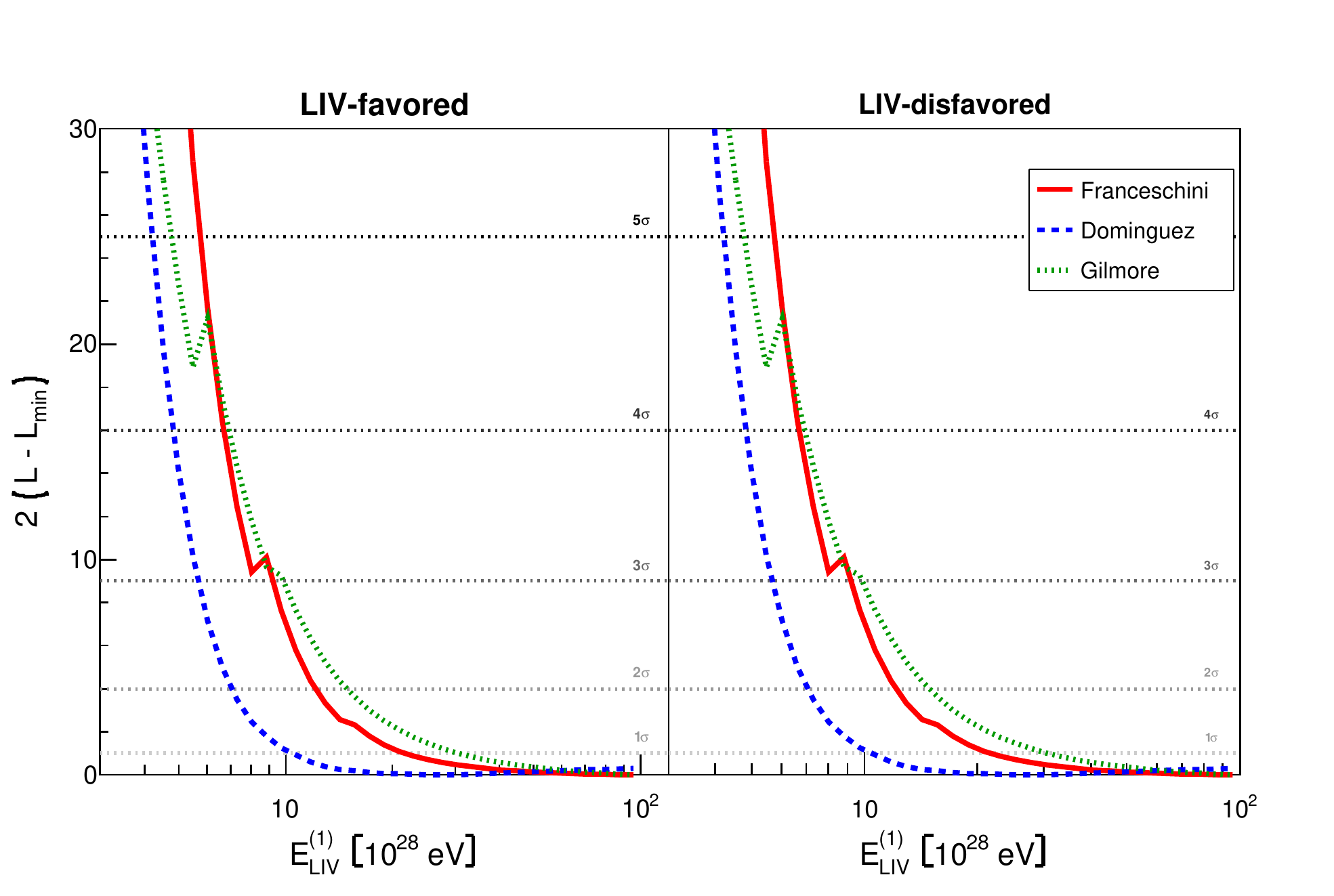}
  \caption{Log-likelihood value as a function of the LIV energy scale for $n=1$. Left plot shows the LIV-favored case and the right plot shows the LIV-disfavored case. Red lines show the results for the Franceschini model, green lines for Gilmore model and blue lines for the Dominguez model. The horizontal dashed lines represent the 1, 2, 3, 4 and 5$\sigma$ confidence levels.}
  \label{fig:EBL}
\end{figure}

The uncertainties in the EBL spectrum are still large~\cite{Biteau:2015,Dominguez}. The reference model (Franceschini~\cite{Franceschini}) was chosen to allow the direct comparison of our results with previous works. At least two other EBL models are also used in the literature: Dominguez~\cite{Dominguez} and Gilmore~\cite{Gilmore}. We repeated our analysis using these two models and the results are shown in Fig.~\ref{fig:EBL} and in Table~\ref{tab:limitsEBL}. The same overall behavior of the log-likelihood curves is obtained and the numerical values of $E_{\mathrm{LIV}}^{(n)}$ at the same confidence level are compatible to the reference analysis.

\begin{table}[h!]
  \centering
  \begin{tabular}{c | c c c | c c c | c c c }
    \hline \hline
    & \multicolumn{3}{|c|}{Franceschini} & \multicolumn{3}{|c|}{Dominguez} & \multicolumn{3}{|c}{Gilmore} \\ \hline
    & $2\sigma$ & $3\sigma$ & $5\sigma$ & $2\sigma$ & $3\sigma$ & $5\sigma$ & $2\sigma$ & $3\sigma$ & $5\sigma$ \\ \hline
    $E_{\mathrm{LIV}}^{(1)} \left[10^{28} \ \mathrm{eV}\right]$ & 12.08 & 9.14 & 5.73 & 6.85 & 5.62 & 4.17 & 14.89 & 9.80 & 4.74 \\
    $E_{\mathrm{LIV}}^{(2)} \left[10^{21} \ \mathrm{eV}\right]$ & 2.38 & 1.69 & 1.42 & 1.56 & 1.40 & 1.14 & 2.17 & 1.78 & 1.31 \\ \hline \hline
  \end{tabular}
  \caption{Limits on the LIV energy scale imposed by this work for different EBL models.}
  \label{tab:limitsEBL}
\end{table}

\subsection{Model of the intrinsic spectrum}
\label{sec:Ab_int}

\begin{figure}[h!]
  \centering
  \includegraphics[width=0.5\textwidth]{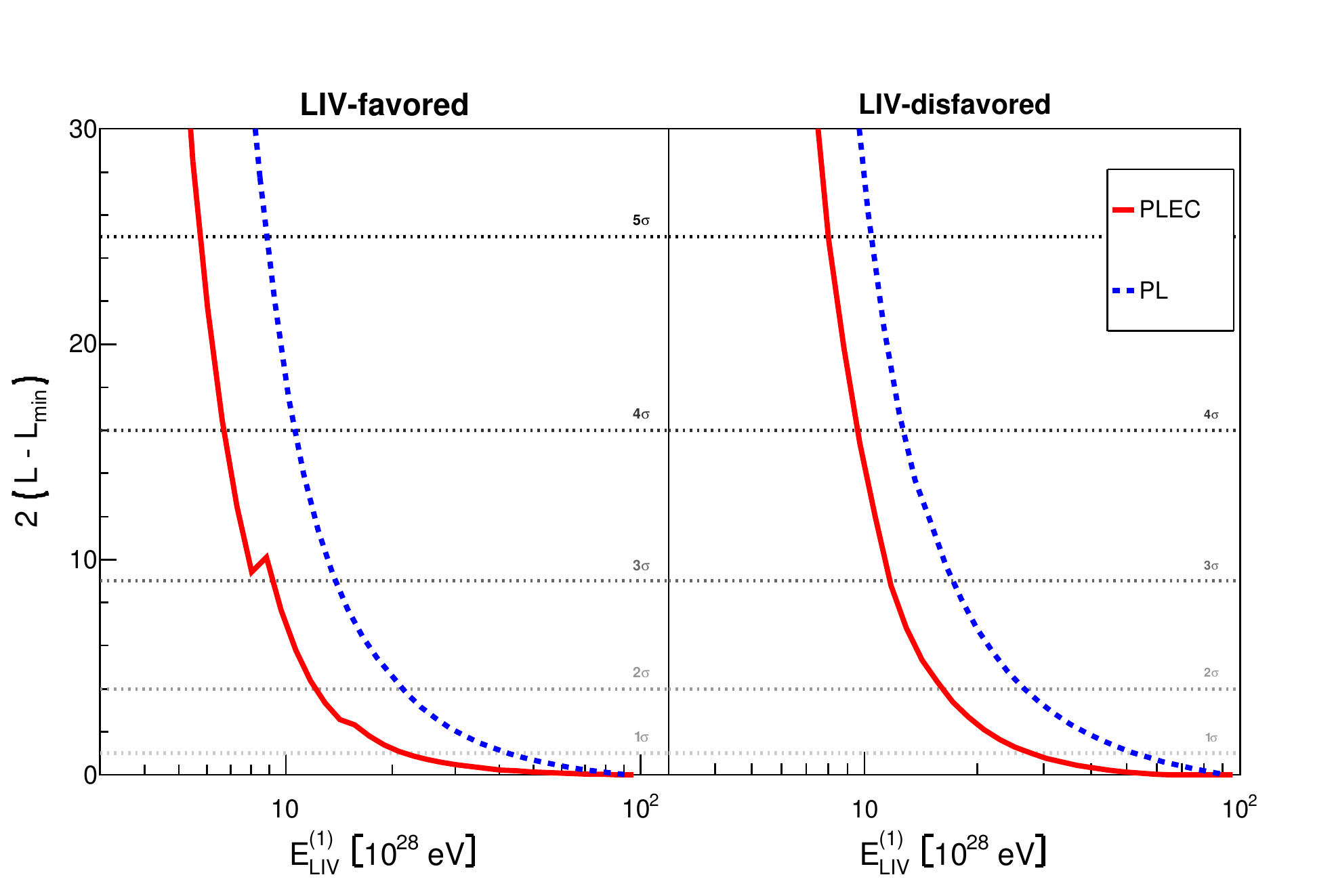}
  \caption{Log-likelihood value as a function of the LIV energy scale for $n=1$. Left plot shows the LIV-favored case and the right plot shows the LIV-disfavored case. The cyan and the red lines show, respectively, the results for the power law and power law with an exponential cutoff parametrizations. The horizontal dashed lines represent the 1, 2, 3, 4 and 5$\sigma$ confidence levels.}
  \label{fig:SpectrumShape}
\end{figure}

\begin{figure}[h!]
  \centering
  \includegraphics[width=0.48\textwidth]{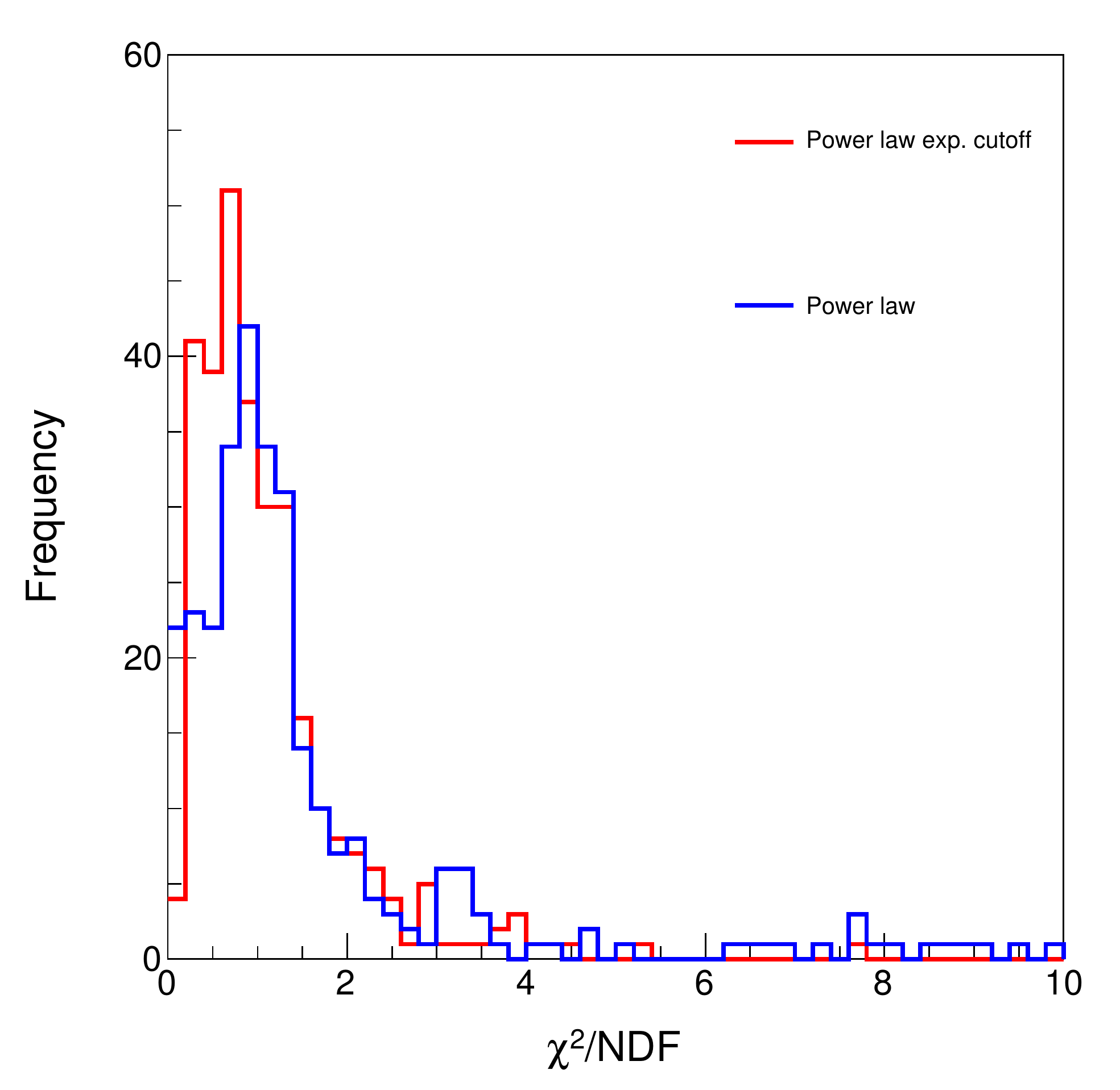}
  \caption{Distribution of $\chi^2$/NDF for the fitted intrinsic spectrum. The blue line represents the distribution obtained using a simple power law parametrization and the red line represents the results using a power law with an exponential cutoff.}
  \label{fig:Chi2}
\end{figure}

We evaluate here the choice of a power law with an exponential cutoff (PLEC) to model the intrinsic spectrum of the sources. This functional shape is motivated by acceleration theory of particles in the source which generate the gamma-ray flux. We test its possible bias by repeating the fit with a simple power law (PL) function. Figure~\ref{fig:SpectrumShape} shows the resulting log-likelihood profile using each parametrization and Fig.~\ref{fig:Chi2} shows the distribution of $\chi^2$/NDF for the reconstructed spectra using both the PL and the PLEC parametrizations.

The spectra are better reconstructed using the PLEC parametrization, with a mean $\chi^2$/NDF of 1.12 and a standard deviation of 0.92, while the PL results in a mean $\chi^2$/NDF of 2.04 with a standard deviation of 3.09. The limits using the PLEC parametrization are also more conservative. The PLEC parametrization contemplates a wider range of possibilities for the bins not used in the reconstruction, including the simple power law. Therefore, using a simple power law for the intrinsic spectra could bias the analysis due to both a bad reconstruction and assuming the behavior of the most energetic bins, where LIV effects emerge.

\subsection{Energy resolution}
\label{sec:Ab_ener}

\begin{figure}[h] 
  \centering
  \includegraphics[width=0.5\textwidth]{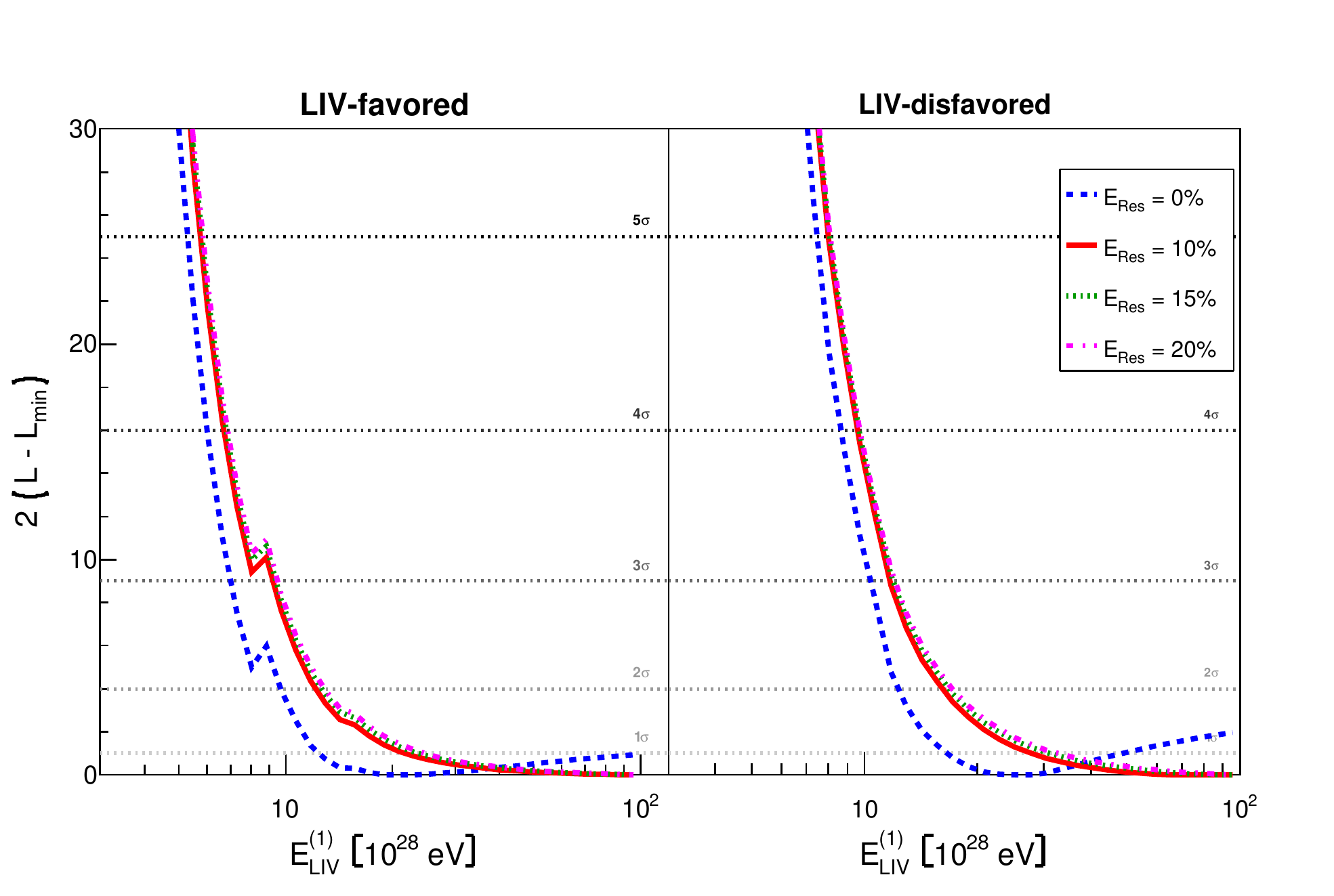}
  \caption{Log-likelihood value as a function of the LIV energy scale for $n=1$. Left plot shows the LIV-favored case and the right plot shows the LIV-disfavored case. The different colored lines show the results for different energy resolutions. The horizontal dashed lines represent the 1, 2, 3, 4 and 5$\sigma$ confidence levels.}
  \label{fig:energy:resolution}
\end{figure}

The reference analysis considers an energy resolution of 10\%. We evaluate the effect of this choice by repeating the analysis with energy resolution of 0\%, 15\% and 20\%. The log-likelihood test is shown in Fig.~\ref{fig:energy:resolution}. The difference is small for energy resolutions of 10\%, 15\% and 20\%. If perfect energy reconstruction is considered (energy resolution 0\%) the analysis bias the results towards a LIV signal. This happens because the bin migration induced by the resolution artificially increases the flux at the highest energies, which mimics the LIV effect.

\subsection{Selection of spectra}
\label{sec:Ab_src}

\begin{figure}[h!]
  \centering
  \includegraphics[width=0.48\textwidth]{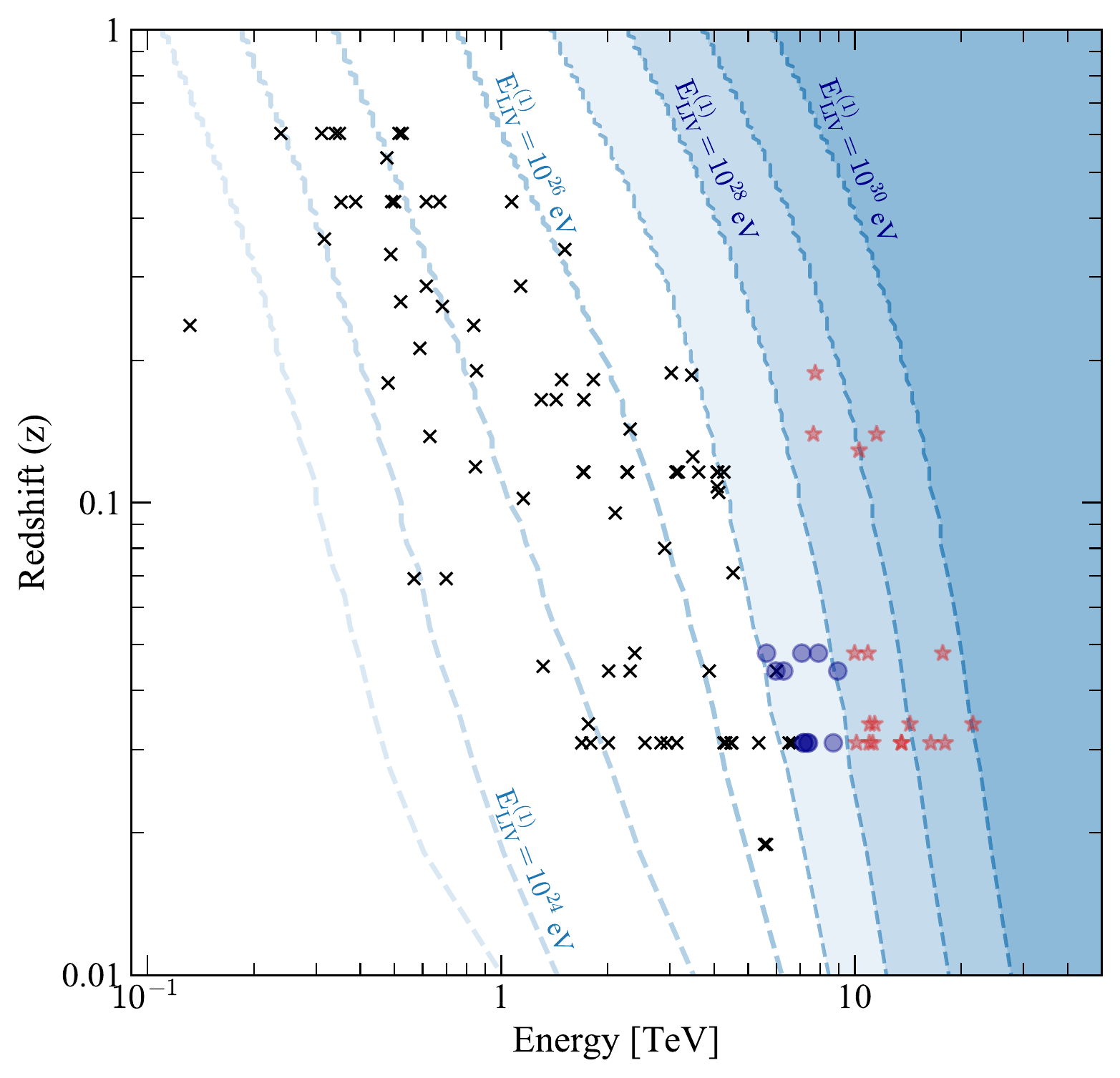}
  \caption{Distance as a function of energy for which $a_{LIV}/a_{LI} > 1.1$ are shown as dashed lines for several $E_{\mathrm{LIV}}^{(1)}$ values. Black crosses, blue circles and the red stars show the distance and $E_{max}$ of all 111 measured energy spectra studied in this work. Reference data set A is composed by the red stars. Data set B analyzed in this section is composed of blue circles and red starts.}
  \label{fig:LIVHorizon_B}
\end{figure}

\begin{figure}[h!]
  \centering
  \includegraphics[width=0.5\textwidth]{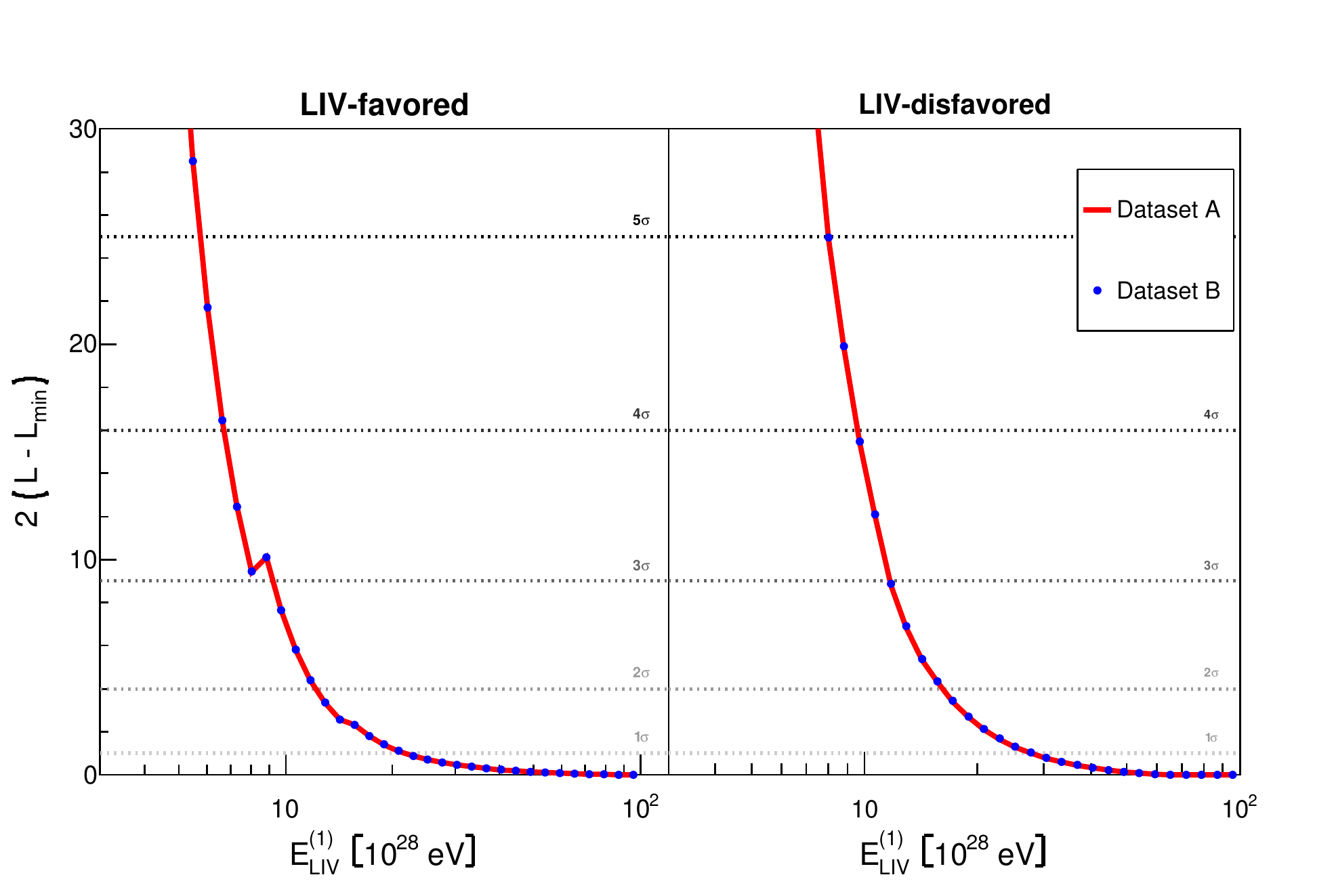}
  \caption{Log-likelihood value as a function of the LIV energy scale for $n=1$. The red and blue lines show the results using data sets A and B, respectively. Both lines overlap. The horizontal dashed lines represent the 1, 2, 3, 4 and 5$\sigma$ confidence levels.}
  \label{fig:dataset}
\end{figure}

The selection procedure presented in the paper chose a subset of 18 measured spectra from 6 sources to be considered in our analysis as is shown in Fig.~\ref{fig:LIVHorizon_B} by the red star. We tested the effect of including other measured spectra in the final result. We arbitrarily included the next 11 spectra from 3 sources in the data analysis shown as blue circles in Fig.~\ref{fig:LIVHorizon_B} and listed in Table IV. The original data set used in~\ref{sec:analysis} is named here data set A and previously listed in Table II. The 29 spectra composed by the original 18 plus 11 arbitrarily added to explore the systematics is named data set B. The corresponding log-likelihood test of the two data sets A and B is shown in Fig.~\ref{fig:dataset}. The curves overlap, we show data set B as blue points for visualization purpose. There was no significant change in the results by adding the extra 11 spectra which proves the efficiency of the selection procedure proposed in the Sec. III of the paper.

\begin{table}[h]
  \centering
  \begin{tabular}{c c c c c}
    \hline \hline
    Source & Redshift & Experiment & Spectrum & {\tiny Reference} \\ \hline
    \multirow{5}{*}{\small Markarian 421} & \multirow{5}{*}{0.031} 
    & \multirow{2}{*}{\small VERITAS} & {\tiny 2006-2008 (highC)} & \cite{Mrk421_VERITAS_low} \\
    & & & {\tiny 2006-2008 (very\_high)} & \cite{Mrk421_VERITAS_low} \\
    & & {\small ARGO-YBJ} & {\tiny 2007-2010  (flux 1)} & \cite{Mrk421_ARGO_flux1} \\
    & & {\small ARGO-YBJ} & {\tiny 2007-2010 (flux 3)} & \cite{Mrk421_ARGO_flux1} \\
    & & {\small ARGO-YBJ} & {\tiny 2007-2010 (flux 4)} & \cite{Mrk421_ARGO_flux1} \\ \hline
    \multirow{3}{*}{\small 1ES 2344+514} & \multirow{3}{*}{0.044} & Whipple & 1995 (b) & \cite{s1ES2344_Whipple_2005_b} \\
    & & \multirow{2}{*}{\small VERITAS} & {\tiny 2007-2008 (low)} & \cite{s1ES2344_VERITAS_low} \\
    & & & 2007-2015 & \cite{s1ES2344_VERITAS_2017} \\ \hline
    \multirow{3}{*}{\small 1ES 1959+650} & \multirow{3}{*}{0.048} & \multirow{3}{*}{\small VERITAS} & 2007-2011 & \cite{s1ES1959_VERITAS} \\
    & & & 2015 & \cite{s1ES1959_VERITAS_2015} \\
    & & & 2016 & \cite{s1ES1959_VERITAS_2015} \\
    \hline \hline
  \end{tabular}
  \caption{List of spectra added in this section to evaluate the influence of the spectra choice in the procedure proposed in the paper. Columns show the name and redshift of the source, the experiment that measured the spectra and a specification of the measured spectrum as in the original publication shown as reference in the last column.}
  \label{tab:sources2}
\end{table}

\subsection{Selection of bins in each measured spectrum}
\label{sec:Ab_bin}

The energy region used to reconstruct the intrinsic spectrum was defined in equation \ref{eq:li:region} and depends on the factor $\rho$. The reference results were obtained using $\rho = 1$ which means we excluded from the intrinsic spectrum reconstruction any energy bin for which the difference between the LI and the LIV attenuation is larger than the error in the measured flux. We evaluate here the systematic effect in the results derived from the choice of $\rho=1$. We repeated the analysis considering $\rho = 3$ and $\rho = 5$. The $\rho$ parameter sets the tolerance for the difference between LI and LIV attenuations. The number of bins used to reconstruct the intrinsic spectrum increases with $\rho$.

\begin{figure}[h]
  \centering
  \includegraphics[width=0.5\textwidth]{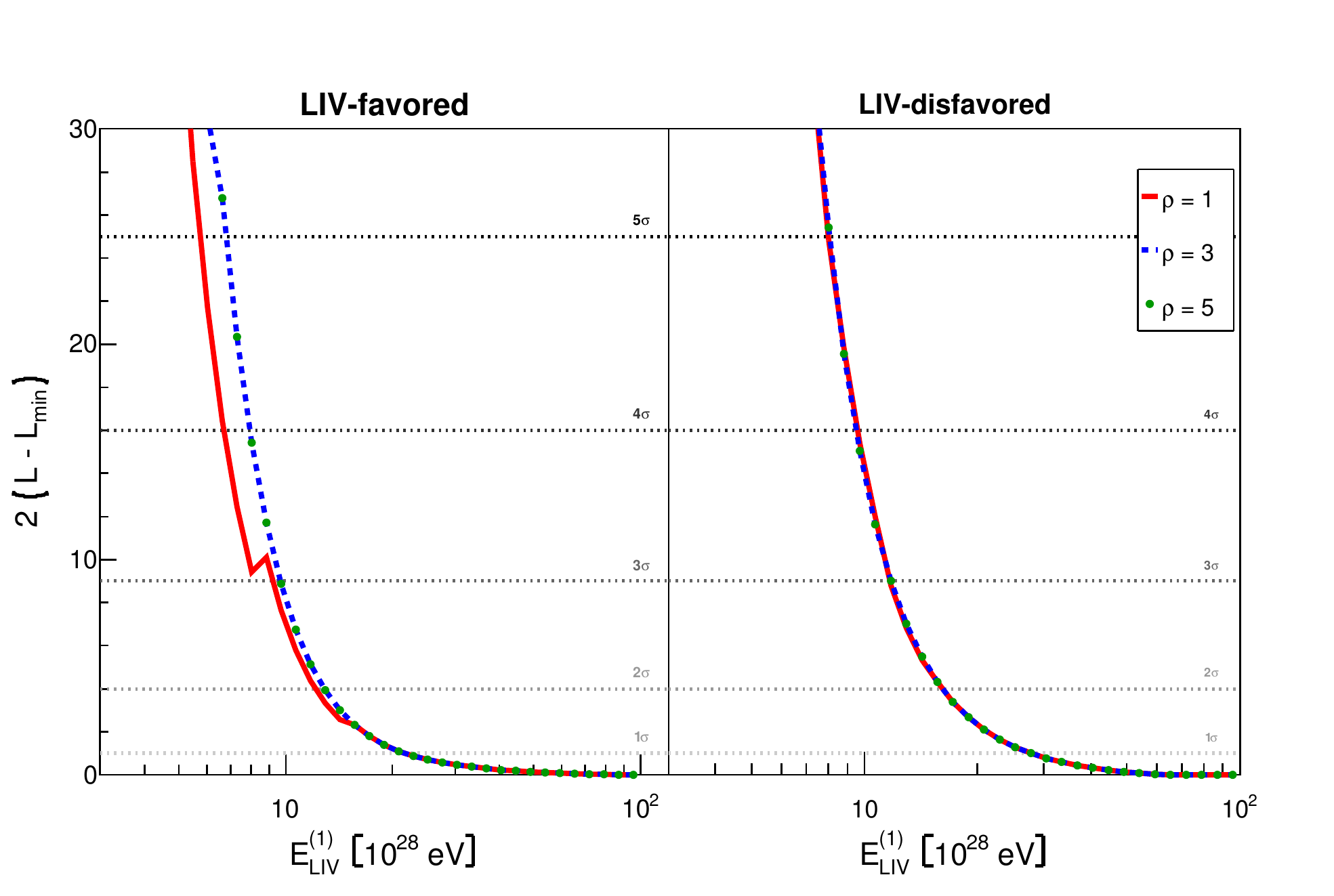}
  \caption{Log-likelihood value as a function of the LIV energy scale for $n=1$. Left plot shows the LIV-favored case and the right plot shows the LIV-disfavored case. The colored lines show the results for different values of $\rho$. The horizontal dashed lines represent the 1, 2, 3, 4 and 5$\sigma$ confidence levels.}
  \label{fig:RFactor}
\end{figure}

Figure~\ref{fig:RFactor} shows the log-likelihood test using $\rho=1$, $\rho=3$ and $\rho=5$.  The test confirms that the overall shape of the log-likelihood curves does not depend on the choice of $\rho$. If the curves overlap, points were used to plot continuous functions for visualization purpose. Most important, this test shows that $\rho=1$ leads to the most conservative LIV limit and that previous analysis which did not take into account this selection might have overestimated the LIV limit.

\newpage

\bibliography{bibfile}

\begin{thebibliography}{67}%
\makeatletter
\providecommand \@ifxundefined [1]{%
 \@ifx{#1\undefined}
}%
\providecommand \@ifnum [1]{%
 \ifnum #1\expandafter \@firstoftwo
 \else \expandafter \@secondoftwo
 \fi
}%
\providecommand \@ifx [1]{%
 \ifx #1\expandafter \@firstoftwo
 \else \expandafter \@secondoftwo
 \fi
}%
\providecommand \natexlab [1]{#1}%
\providecommand \enquote  [1]{``#1''}%
\providecommand \bibnamefont  [1]{#1}%
\providecommand \bibfnamefont [1]{#1}%
\providecommand \citenamefont [1]{#1}%
\providecommand \href@noop [0]{\@secondoftwo}%
\providecommand \href [0]{\begingroup \@sanitize@url \@href}%
\providecommand \@href[1]{\@@startlink{#1}\@@href}%
\providecommand \@@href[1]{\endgroup#1\@@endlink}%
\providecommand \@sanitize@url [0]{\catcode `\\12\catcode `\$12\catcode
  `\&12\catcode `\#12\catcode `\^12\catcode `\_12\catcode `\%12\relax}%
\providecommand \@@startlink[1]{}%
\providecommand \@@endlink[0]{}%
\providecommand \url  [0]{\begingroup\@sanitize@url \@url }%
\providecommand \@url [1]{\endgroup\@href {#1}{\urlprefix }}%
\providecommand \urlprefix  [0]{URL }%
\providecommand \Eprint [0]{\href }%
\providecommand \doibase [0]{http://dx.doi.org/}%
\providecommand \selectlanguage [0]{\@gobble}%
\providecommand \bibinfo  [0]{\@secondoftwo}%
\providecommand \bibfield  [0]{\@secondoftwo}%
\providecommand \translation [1]{[#1]}%
\providecommand \BibitemOpen [0]{}%
\providecommand \bibitemStop [0]{}%
\providecommand \bibitemNoStop [0]{.\EOS\space}%
\providecommand \EOS [0]{\spacefactor3000\relax}%
\providecommand \BibitemShut  [1]{\csname bibitem#1\endcsname}%
\let\auto@bib@innerbib\@empty
\bibitem [{\citenamefont {Nambu}(1968)}]{NAMBU}%
  \BibitemOpen
  \bibfield  {author} {\bibinfo {author} {\bibfnamefont {Y.}~\bibnamefont
  {Nambu}},\ }\href@noop {} {\bibfield  {journal} {\bibinfo  {journal}
  {Supplement of the Progress of Theoretical Physics}\ }\textbf {\bibinfo
  {volume} {E68}},\ \bibinfo {pages} {190} (\bibinfo {year}
  {1968})}\BibitemShut {NoStop}%
\bibitem [{\citenamefont {Kostelecky}\ and\ \citenamefont
  {Samuel}(1989)}]{Kostelecky:1988zi}%
  \BibitemOpen
  \bibfield  {author} {\bibinfo {author} {\bibfnamefont {V.~A.}\ \bibnamefont
  {Kostelecky}}\ and\ \bibinfo {author} {\bibfnamefont {S.}~\bibnamefont
  {Samuel}},\ }\href {\doibase 10.1103/PhysRevD.39.683} {\bibfield  {journal}
  {\bibinfo  {journal} {Phys. Rev.}\ }\textbf {\bibinfo {volume} {D39}},\
  \bibinfo {pages} {683} (\bibinfo {year} {1989})}\BibitemShut {NoStop}%
\bibitem [{\citenamefont {Colladay}\ and\ \citenamefont
  {Kostelecky}(1998)}]{Colladay:1998fq}%
  \BibitemOpen
  \bibfield  {author} {\bibinfo {author} {\bibfnamefont {D.}~\bibnamefont
  {Colladay}}\ and\ \bibinfo {author} {\bibfnamefont {V.~A.}\ \bibnamefont
  {Kostelecky}},\ }\href {\doibase 10.1103/PhysRevD.58.116002} {\bibfield
  {journal} {\bibinfo  {journal} {Phys. Rev.}\ }\textbf {\bibinfo {volume}
  {D58}},\ \bibinfo {pages} {116002} (\bibinfo {year} {1998})},\ \Eprint
  {http://arxiv.org/abs/hep-ph/9809521} {arXiv:hep-ph/9809521 [hep-ph]}
  \BibitemShut {NoStop}%
\bibitem [{\citenamefont {Ellis}\ \emph {et~al.}(1999)\citenamefont {Ellis},
  \citenamefont {Mavromatos},\ and\ \citenamefont {Nanopoulos}}]{QG4}%
  \BibitemOpen
  \bibfield  {author} {\bibinfo {author} {\bibfnamefont {J.}~\bibnamefont
  {Ellis}}, \bibinfo {author} {\bibfnamefont {N.~E.}\ \bibnamefont
  {Mavromatos}}, \ and\ \bibinfo {author} {\bibfnamefont {D.~V.}\ \bibnamefont
  {Nanopoulos}},\ }\href@noop {} {\bibfield  {journal} {\bibinfo  {journal}
  {Phys. Rev. D}\ }\textbf {\bibinfo {volume} {61}},\ \bibinfo {pages} {027503}
  (\bibinfo {year} {1999})}\BibitemShut {NoStop}%
\bibitem [{\citenamefont {Gambini}\ and\ \citenamefont {Pullin}(1999)}]{QG5}%
  \BibitemOpen
  \bibfield  {author} {\bibinfo {author} {\bibfnamefont {R.}~\bibnamefont
  {Gambini}}\ and\ \bibinfo {author} {\bibfnamefont {J.}~\bibnamefont
  {Pullin}},\ }\href@noop {} {\bibfield  {journal} {\bibinfo  {journal} {Phys.
  Rev.}\ }\textbf {\bibinfo {volume} {59}},\ \bibinfo {pages} {124021}
  (\bibinfo {year} {1999})}\BibitemShut {NoStop}%
\bibitem [{\citenamefont {Amelino-Camelia}(2001)}]{QG1}%
  \BibitemOpen
  \bibfield  {author} {\bibinfo {author} {\bibfnamefont {G.}~\bibnamefont
  {Amelino-Camelia}},\ }\href@noop {} {\bibfield  {journal} {\bibinfo
  {journal} {Nature}\ }\textbf {\bibinfo {volume} {410}},\ \bibinfo {pages}
  {1065} (\bibinfo {year} {2001})}\BibitemShut {NoStop}%
\bibitem [{\citenamefont {Alfaro}(2005)}]{ALFARO}%
  \BibitemOpen
  \bibfield  {author} {\bibinfo {author} {\bibfnamefont {J.}~\bibnamefont
  {Alfaro}},\ }\href@noop {} {\bibfield  {journal} {\bibinfo  {journal} {Phys.
  Rev. Lett.}\ }\textbf {\bibinfo {volume} {94}},\ \bibinfo {pages} {221302}
  (\bibinfo {year} {2005})}\BibitemShut {NoStop}%
\bibitem [{\citenamefont {Potting}(2013)}]{Pot}%
  \BibitemOpen
  \bibfield  {author} {\bibinfo {author} {\bibfnamefont {R.}~\bibnamefont
  {Potting}},\ }\href@noop {} {\bibfield  {journal} {\bibinfo  {journal}
  {Journal of Physics: Conference Series}\ }\textbf {\bibinfo {volume} {Volume
  447}},\ \bibinfo {pages} {012009} (\bibinfo {year} {2013})}\BibitemShut
  {NoStop}%
\bibitem [{\citenamefont {Audren}\ \emph {et~al.}(2013)\citenamefont {Audren},
  \citenamefont {Blas}, \citenamefont {Lesgourgues},\ and\ \citenamefont
  {Sibiryakov}}]{Audren:2013dwa}%
  \BibitemOpen
  \bibfield  {author} {\bibinfo {author} {\bibfnamefont {B.}~\bibnamefont
  {Audren}}, \bibinfo {author} {\bibfnamefont {D.}~\bibnamefont {Blas}},
  \bibinfo {author} {\bibfnamefont {J.}~\bibnamefont {Lesgourgues}}, \ and\
  \bibinfo {author} {\bibfnamefont {S.}~\bibnamefont {Sibiryakov}},\ }\href
  {\doibase 10.1088/1475-7516/2013/08/039} {\bibfield  {journal} {\bibinfo
  {journal} {JCAP}\ }\textbf {\bibinfo {volume} {1308}},\ \bibinfo {pages}
  {039} (\bibinfo {year} {2013})},\ \Eprint {http://arxiv.org/abs/1305.0009}
  {arXiv:1305.0009 [astro-ph.CO]} \BibitemShut {NoStop}%
\bibitem [{\citenamefont {Bluhm}(2014)}]{Bluhm}%
  \BibitemOpen
  \bibfield  {author} {\bibinfo {author} {\bibfnamefont {R.}~\bibnamefont
  {Bluhm}},\ }\enquote {\bibinfo {title} {Springer handbook of spacetime},}\ \
  (\bibinfo {year} {2014})\ Chap.\ \bibinfo {chapter} {Observational
  Constraints on Local Lorentz Invariance}, pp.\ \bibinfo {pages}
  {485--507}\BibitemShut {NoStop}%
\bibitem [{\citenamefont {Calcagni}(2017)}]{Calcagni:2016zqv}%
  \BibitemOpen
  \bibfield  {author} {\bibinfo {author} {\bibfnamefont {G.}~\bibnamefont
  {Calcagni}},\ }\href {\doibase 10.1140/epjc/s10052-017-4841-6} {\bibfield
  {journal} {\bibinfo  {journal} {Eur. Phys. J.}\ }\textbf {\bibinfo {volume}
  {C77}},\ \bibinfo {pages} {291} (\bibinfo {year} {2017})},\ \Eprint
  {http://arxiv.org/abs/1603.03046} {arXiv:1603.03046 [gr-qc]} \BibitemShut
  {NoStop}%
\bibitem [{\citenamefont {Bettoni}\ \emph {et~al.}(2017)\citenamefont
  {Bettoni}, \citenamefont {Nusser}, \citenamefont {Blas},\ and\ \citenamefont
  {Sibiryakov}}]{Bettoni:2017lxf}%
  \BibitemOpen
  \bibfield  {author} {\bibinfo {author} {\bibfnamefont {D.}~\bibnamefont
  {Bettoni}}, \bibinfo {author} {\bibfnamefont {A.}~\bibnamefont {Nusser}},
  \bibinfo {author} {\bibfnamefont {D.}~\bibnamefont {Blas}}, \ and\ \bibinfo
  {author} {\bibfnamefont {S.}~\bibnamefont {Sibiryakov}},\ }\href {\doibase
  10.1088/1475-7516/2017/05/024} {\bibfield  {journal} {\bibinfo  {journal}
  {JCAP}\ }\textbf {\bibinfo {volume} {1705}},\ \bibinfo {pages} {024}
  (\bibinfo {year} {2017})},\ \Eprint {http://arxiv.org/abs/1702.07726}
  {arXiv:1702.07726 [astro-ph.CO]} \BibitemShut {NoStop}%
\bibitem [{\citenamefont {Liberati}\ and\ \citenamefont
  {Maccione}(2009)}]{bib:liv:tests:astropart}%
  \BibitemOpen
  \bibfield  {author} {\bibinfo {author} {\bibfnamefont {S.}~\bibnamefont
  {Liberati}}\ and\ \bibinfo {author} {\bibfnamefont {L.}~\bibnamefont
  {Maccione}},\ }\href {\doibase 10.1146/annurev.nucl.010909.083640} {\bibfield
   {journal} {\bibinfo  {journal} {Annual Review of Nuclear and Particle
  Science}\ }\textbf {\bibinfo {volume} {59}},\ \bibinfo {pages} {245}
  (\bibinfo {year} {2009})}\BibitemShut {NoStop}%
\bibitem [{\citenamefont {Coleman}\ and\ \citenamefont
  {Glashow}(1997)}]{Coleman:1997xq}%
  \BibitemOpen
  \bibfield  {author} {\bibinfo {author} {\bibfnamefont {S.~R.}\ \bibnamefont
  {Coleman}}\ and\ \bibinfo {author} {\bibfnamefont {S.~L.}\ \bibnamefont
  {Glashow}},\ }\href {\doibase 10.1016/S0370-2693(97)00638-2} {\bibfield
  {journal} {\bibinfo  {journal} {Phys. Lett.}\ }\textbf {\bibinfo {volume}
  {B405}},\ \bibinfo {pages} {249} (\bibinfo {year} {1997})},\ \Eprint
  {http://arxiv.org/abs/hep-ph/9703240} {arXiv:hep-ph/9703240 [hep-ph]}
  \BibitemShut {NoStop}%
\bibitem [{\citenamefont {Amelino-Camelia}\ \emph {et~al.}(1998)\citenamefont
  {Amelino-Camelia}, \citenamefont {Ellis}, \citenamefont {Mavromatos},
  \citenamefont {Nanopoulos},\ and\ \citenamefont
  {Sarkar}}]{AmelinoEllis:1998}%
  \BibitemOpen
  \bibfield  {author} {\bibinfo {author} {\bibfnamefont {G.}~\bibnamefont
  {Amelino-Camelia}}, \bibinfo {author} {\bibfnamefont {J.~R.}\ \bibnamefont
  {Ellis}}, \bibinfo {author} {\bibfnamefont {N.~E.}\ \bibnamefont
  {Mavromatos}}, \bibinfo {author} {\bibfnamefont {D.~V.}\ \bibnamefont
  {Nanopoulos}}, \ and\ \bibinfo {author} {\bibfnamefont {S.}~\bibnamefont
  {Sarkar}},\ }\href {\doibase 10.1038/31647} {\bibfield  {journal} {\bibinfo
  {journal} {Nature}\ }\textbf {\bibinfo {volume} {393}},\ \bibinfo {pages}
  {763} (\bibinfo {year} {1998})},\ \Eprint
  {http://arxiv.org/abs/astro-ph/9712103} {arXiv:astro-ph/9712103 [astro-ph]}
  \BibitemShut {NoStop}%
\bibitem [{\citenamefont {Coleman}\ and\ \citenamefont
  {Glashow}(1999)}]{Coleman:1998ti}%
  \BibitemOpen
  \bibfield  {author} {\bibinfo {author} {\bibfnamefont {S.~R.}\ \bibnamefont
  {Coleman}}\ and\ \bibinfo {author} {\bibfnamefont {S.~L.}\ \bibnamefont
  {Glashow}},\ }\href {\doibase 10.1103/PhysRevD.59.116008} {\bibfield
  {journal} {\bibinfo  {journal} {Phys. Rev.}\ }\textbf {\bibinfo {volume}
  {D59}},\ \bibinfo {pages} {116008} (\bibinfo {year} {1999})},\ \Eprint
  {http://arxiv.org/abs/hep-ph/9812418} {arXiv:hep-ph/9812418 [hep-ph]}
  \BibitemShut {NoStop}%
\bibitem [{\citenamefont {Stecker}\ and\ \citenamefont
  {Glashow}(2001)}]{Stecker:2001vb}%
  \BibitemOpen
  \bibfield  {author} {\bibinfo {author} {\bibfnamefont {F.~W.}\ \bibnamefont
  {Stecker}}\ and\ \bibinfo {author} {\bibfnamefont {S.~L.}\ \bibnamefont
  {Glashow}},\ }\href {\doibase 10.1016/S0927-6505(01)00137-2} {\bibfield
  {journal} {\bibinfo  {journal} {Astropart. Phys.}\ }\textbf {\bibinfo
  {volume} {16}},\ \bibinfo {pages} {97} (\bibinfo {year} {2001})},\ \Eprint
  {http://arxiv.org/abs/astro-ph/0102226} {arXiv:astro-ph/0102226 [astro-ph]}
  \BibitemShut {NoStop}%
\bibitem [{\citenamefont {Stecker}(2003)}]{Stecker:2003pw}%
  \BibitemOpen
  \bibfield  {author} {\bibinfo {author} {\bibfnamefont {F.~W.}\ \bibnamefont
  {Stecker}},\ }\href {\doibase 10.1016/j.astropartphys.2003.08.006} {\bibfield
   {journal} {\bibinfo  {journal} {Astropart. Phys.}\ }\textbf {\bibinfo
  {volume} {20}},\ \bibinfo {pages} {85} (\bibinfo {year} {2003})},\ \Eprint
  {http://arxiv.org/abs/astro-ph/0308214} {arXiv:astro-ph/0308214 [astro-ph]}
  \BibitemShut {NoStop}%
\bibitem [{\citenamefont {Jacobson}\ \emph {et~al.}(2003)\citenamefont
  {Jacobson}, \citenamefont {Liberati},\ and\ \citenamefont
  {Mattingly}}]{Jacobson:2002}%
  \BibitemOpen
  \bibfield  {author} {\bibinfo {author} {\bibfnamefont {T.}~\bibnamefont
  {Jacobson}}, \bibinfo {author} {\bibfnamefont {S.}~\bibnamefont {Liberati}},
  \ and\ \bibinfo {author} {\bibfnamefont {D.}~\bibnamefont {Mattingly}},\
  }\href {\doibase 10.1103/PhysRevD.67.124011} {\bibfield  {journal} {\bibinfo
  {journal} {Phys. Rev.}\ }\textbf {\bibinfo {volume} {D67}},\ \bibinfo {pages}
  {124011} (\bibinfo {year} {2003})},\ \Eprint
  {http://arxiv.org/abs/hep-ph/0209264} {arXiv:hep-ph/0209264 [hep-ph]}
  \BibitemShut {NoStop}%
\bibitem [{\citenamefont {Stecker}\ and\ \citenamefont
  {Scully}(2005)}]{Stecker:2004}%
  \BibitemOpen
  \bibfield  {author} {\bibinfo {author} {\bibfnamefont {F.~W.}\ \bibnamefont
  {Stecker}}\ and\ \bibinfo {author} {\bibfnamefont {S.~T.}\ \bibnamefont
  {Scully}},\ }\href {\doibase 10.1016/j.astropartphys.2005.01.001} {\bibfield
  {journal} {\bibinfo  {journal} {Astropart. Phys.}\ }\textbf {\bibinfo
  {volume} {23}},\ \bibinfo {pages} {203} (\bibinfo {year} {2005})},\ \Eprint
  {http://arxiv.org/abs/astro-ph/0412495} {arXiv:astro-ph/0412495 [astro-ph]}
  \BibitemShut {NoStop}%
\bibitem [{\citenamefont {Ellis}\ \emph {et~al.}(2006)\citenamefont {Ellis},
  \citenamefont {Mavromatos}, \citenamefont {Nanopoulos}, \citenamefont
  {Sakharov},\ and\ \citenamefont {Sarkisyan}}]{Ellis2006402}%
  \BibitemOpen
  \bibfield  {author} {\bibinfo {author} {\bibfnamefont {J.~R.}\ \bibnamefont
  {Ellis}}, \bibinfo {author} {\bibfnamefont {N.~E.}\ \bibnamefont
  {Mavromatos}}, \bibinfo {author} {\bibfnamefont {D.~V.}\ \bibnamefont
  {Nanopoulos}}, \bibinfo {author} {\bibfnamefont {A.~S.}\ \bibnamefont
  {Sakharov}}, \ and\ \bibinfo {author} {\bibfnamefont {E.~K.~G.}\ \bibnamefont
  {Sarkisyan}},\ }\href {\doibase 10.1016/j.astropartphys.2006.04.001,
  10.1016/j.astropartphys.2007.12.003} {\bibfield  {journal} {\bibinfo
  {journal} {Astropart. Phys.}\ }\textbf {\bibinfo {volume} {25}},\ \bibinfo
  {pages} {402} (\bibinfo {year} {2006})},\ \bibinfo {note} {[Erratum:
  Astropart. Phys.29,158(2008)]},\ \Eprint {http://arxiv.org/abs/0712.2781}
  {arXiv:0712.2781 [astro-ph]} \BibitemShut {NoStop}%
\bibitem [{\citenamefont {Galaverni}\ and\ \citenamefont
  {Sigl}(2008{\natexlab{a}})}]{Gunter:2007}%
  \BibitemOpen
  \bibfield  {author} {\bibinfo {author} {\bibfnamefont {M.}~\bibnamefont
  {Galaverni}}\ and\ \bibinfo {author} {\bibfnamefont {G.}~\bibnamefont
  {Sigl}},\ }\href {\doibase 10.1103/PhysRevLett.100.021102} {\bibfield
  {journal} {\bibinfo  {journal} {Phys. Rev. Lett.}\ }\textbf {\bibinfo
  {volume} {100}},\ \bibinfo {pages} {021102} (\bibinfo {year}
  {2008}{\natexlab{a}})},\ \Eprint {http://arxiv.org/abs/0708.1737}
  {arXiv:0708.1737 [astro-ph]} \BibitemShut {NoStop}%
\bibitem [{\citenamefont {Galaverni}\ and\ \citenamefont
  {Sigl}(2008{\natexlab{b}})}]{Gunter:2008}%
  \BibitemOpen
  \bibfield  {author} {\bibinfo {author} {\bibfnamefont {M.}~\bibnamefont
  {Galaverni}}\ and\ \bibinfo {author} {\bibfnamefont {G.}~\bibnamefont
  {Sigl}},\ }\href {\doibase 10.1103/PhysRevD.78.063003} {\bibfield  {journal}
  {\bibinfo  {journal} {Phys. Rev.}\ }\textbf {\bibinfo {volume} {D78}},\
  \bibinfo {pages} {063003} (\bibinfo {year} {2008}{\natexlab{b}})},\ \Eprint
  {http://arxiv.org/abs/0807.1210} {arXiv:0807.1210 [astro-ph]} \BibitemShut
  {NoStop}%
\bibitem [{\citenamefont {Albert}\ \emph {et~al.}(2008)\citenamefont {Albert}
  \emph {et~al.}}]{ALBERT2008253}%
  \BibitemOpen
  \bibfield  {author} {\bibinfo {author} {\bibfnamefont {J.}~\bibnamefont
  {Albert}} \emph {et~al.} (\bibinfo {collaboration} {MAGIC, Other
  Contributors}),\ }\href {\doibase 10.1016/j.physletb.2008.08.053} {\bibfield
  {journal} {\bibinfo  {journal} {Phys. Lett.}\ }\textbf {\bibinfo {volume}
  {B668}},\ \bibinfo {pages} {253} (\bibinfo {year} {2008})},\ \Eprint
  {http://arxiv.org/abs/0708.2889} {arXiv:0708.2889 [astro-ph]} \BibitemShut
  {NoStop}%
\bibitem [{\citenamefont {Stecker}\ and\ \citenamefont
  {Scully}(2009)}]{Stecker:2009}%
  \BibitemOpen
  \bibfield  {author} {\bibinfo {author} {\bibfnamefont {F.~W.}\ \bibnamefont
  {Stecker}}\ and\ \bibinfo {author} {\bibfnamefont {S.~T.}\ \bibnamefont
  {Scully}},\ }\href {\doibase 10.1088/1367-2630/11/8/085003} {\bibfield
  {journal} {\bibinfo  {journal} {New J. Phys.}\ }\textbf {\bibinfo {volume}
  {11}},\ \bibinfo {pages} {085003} (\bibinfo {year} {2009})},\ \Eprint
  {http://arxiv.org/abs/0906.1735} {arXiv:0906.1735 [astro-ph.HE]} \BibitemShut
  {NoStop}%
\bibitem [{\citenamefont {Xu}\ and\ \citenamefont {Ma}(2016)}]{XU201672}%
  \BibitemOpen
  \bibfield  {author} {\bibinfo {author} {\bibfnamefont {H.}~\bibnamefont
  {Xu}}\ and\ \bibinfo {author} {\bibfnamefont {B.-Q.}\ \bibnamefont {Ma}},\
  }\href {\doibase 10.1016/j.astropartphys.2016.05.008} {\bibfield  {journal}
  {\bibinfo  {journal} {Astropart. Phys.}\ }\textbf {\bibinfo {volume} {82}},\
  \bibinfo {pages} {72} (\bibinfo {year} {2016})},\ \Eprint
  {http://arxiv.org/abs/1607.03203} {arXiv:1607.03203 [hep-ph]} \BibitemShut
  {NoStop}%
\bibitem [{\citenamefont {Chang}\ \emph {et~al.}(2016)\citenamefont {Chang},
  \citenamefont {Li}, \citenamefont {Lin}, \citenamefont {Sang}, \citenamefont
  {Wang},\ and\ \citenamefont {Wang}}]{1674-1137-40-4-045102}%
  \BibitemOpen
  \bibfield  {author} {\bibinfo {author} {\bibfnamefont {Z.}~\bibnamefont
  {Chang}}, \bibinfo {author} {\bibfnamefont {X.}~\bibnamefont {Li}}, \bibinfo
  {author} {\bibfnamefont {H.-N.}\ \bibnamefont {Lin}}, \bibinfo {author}
  {\bibfnamefont {Y.}~\bibnamefont {Sang}}, \bibinfo {author} {\bibfnamefont
  {P.}~\bibnamefont {Wang}}, \ and\ \bibinfo {author} {\bibfnamefont
  {S.}~\bibnamefont {Wang}},\ }\href {\doibase 10.1088/1674-1137/40/4/045102}
  {\bibfield  {journal} {\bibinfo  {journal} {Chin. Phys.}\ }\textbf {\bibinfo
  {volume} {C40}},\ \bibinfo {pages} {045102} (\bibinfo {year} {2016})},\
  \Eprint {http://arxiv.org/abs/1506.08495} {arXiv:1506.08495 [astro-ph.HE]}
  \BibitemShut {NoStop}%
\bibitem [{\citenamefont {Ellis}\ and\ \citenamefont
  {Mavromatos}(2013)}]{ELLIS201350}%
  \BibitemOpen
  \bibfield  {author} {\bibinfo {author} {\bibfnamefont {J.}~\bibnamefont
  {Ellis}}\ and\ \bibinfo {author} {\bibfnamefont {N.~E.}\ \bibnamefont
  {Mavromatos}},\ }\href {\doibase 10.1016/j.astropartphys.2012.05.004}
  {\bibfield  {journal} {\bibinfo  {journal} {Astropart. Phys.}\ }\textbf
  {\bibinfo {volume} {43}},\ \bibinfo {pages} {50} (\bibinfo {year} {2013})},\
  \Eprint {http://arxiv.org/abs/1111.1178} {arXiv:1111.1178 [astro-ph.HE]}
  \BibitemShut {NoStop}%
\bibitem [{\citenamefont {Fairbairn}\ \emph {et~al.}(2014)\citenamefont
  {Fairbairn}, \citenamefont {Nilsson}, \citenamefont {Ellis}, \citenamefont
  {Hinton},\ and\ \citenamefont {White}}]{Farbairn:2014}%
  \BibitemOpen
  \bibfield  {author} {\bibinfo {author} {\bibfnamefont {M.}~\bibnamefont
  {Fairbairn}}, \bibinfo {author} {\bibfnamefont {A.}~\bibnamefont {Nilsson}},
  \bibinfo {author} {\bibfnamefont {J.}~\bibnamefont {Ellis}}, \bibinfo
  {author} {\bibfnamefont {J.}~\bibnamefont {Hinton}}, \ and\ \bibinfo {author}
  {\bibfnamefont {R.}~\bibnamefont {White}},\ }\href {\doibase
  10.1088/1475-7516/2014/06/005} {\bibfield  {journal} {\bibinfo  {journal}
  {JCAP}\ }\textbf {\bibinfo {volume} {1406}},\ \bibinfo {pages} {005}
  (\bibinfo {year} {2014})},\ \Eprint {http://arxiv.org/abs/1401.8178}
  {arXiv:1401.8178 [astro-ph.HE]} \BibitemShut {NoStop}%
\bibitem [{\citenamefont {Tavecchio}\ and\ \citenamefont
  {Bonnoli}(2016)}]{Tavecchio:2015}%
  \BibitemOpen
  \bibfield  {author} {\bibinfo {author} {\bibfnamefont {F.}~\bibnamefont
  {Tavecchio}}\ and\ \bibinfo {author} {\bibfnamefont {G.}~\bibnamefont
  {Bonnoli}},\ }\href {\doibase 10.1051/0004-6361/201526071} {\bibfield
  {journal} {\bibinfo  {journal} {Astron. Astrophys.}\ }\textbf {\bibinfo
  {volume} {585}},\ \bibinfo {pages} {A25} (\bibinfo {year} {2016})},\ \Eprint
  {http://arxiv.org/abs/1510.00980} {arXiv:1510.00980 [astro-ph.HE]}
  \BibitemShut {NoStop}%
\bibitem [{\citenamefont {Biteau}\ and\ \citenamefont
  {Williams}(2015)}]{Biteau:2015}%
  \BibitemOpen
  \bibfield  {author} {\bibinfo {author} {\bibfnamefont {J.}~\bibnamefont
  {Biteau}}\ and\ \bibinfo {author} {\bibfnamefont {D.~A.}\ \bibnamefont
  {Williams}},\ }\href {\doibase 10.1088/0004-637X/812/1/60} {\bibfield
  {journal} {\bibinfo  {journal} {Astrophys. J.}\ }\textbf {\bibinfo {volume}
  {812}},\ \bibinfo {pages} {60} (\bibinfo {year} {2015})},\ \Eprint
  {http://arxiv.org/abs/1502.04166} {arXiv:1502.04166 [astro-ph.CO]}
  \BibitemShut {NoStop}%
\bibitem [{\citenamefont {Mart\'inez-Huerta}\ and\ \citenamefont
  {P\'erez-Lorenzana}(2017)}]{Martinez-Huerta:2016azo}%
  \BibitemOpen
  \bibfield  {author} {\bibinfo {author} {\bibfnamefont {H.}~\bibnamefont
  {Mart\'inez-Huerta}}\ and\ \bibinfo {author} {\bibfnamefont {A.}~\bibnamefont
  {P\'erez-Lorenzana}},\ }\href {\doibase 10.1103/PhysRevD.95.063001}
  {\bibfield  {journal} {\bibinfo  {journal} {Phys. Rev.}\ }\textbf {\bibinfo
  {volume} {D95}},\ \bibinfo {pages} {063001} (\bibinfo {year} {2017})},\
  \Eprint {http://arxiv.org/abs/1610.00047} {arXiv:1610.00047 [astro-ph.HE]}
  \BibitemShut {NoStop}%
\bibitem [{\citenamefont {Rubtsov}\ \emph {et~al.}(2017)\citenamefont
  {Rubtsov}, \citenamefont {Satunin},\ and\ \citenamefont
  {Sibiryakov}}]{Rubtosov:2017}%
  \BibitemOpen
  \bibfield  {author} {\bibinfo {author} {\bibfnamefont {G.}~\bibnamefont
  {Rubtsov}}, \bibinfo {author} {\bibfnamefont {P.}~\bibnamefont {Satunin}}, \
  and\ \bibinfo {author} {\bibfnamefont {S.}~\bibnamefont {Sibiryakov}},\
  }\href {\doibase 10.1088/1475-7516/2017/05/049} {\bibfield  {journal}
  {\bibinfo  {journal} {JCAP}\ }\textbf {\bibinfo {volume} {1705}},\ \bibinfo
  {pages} {049} (\bibinfo {year} {2017})},\ \Eprint
  {http://arxiv.org/abs/1611.10125} {arXiv:1611.10125 [astro-ph.HE]}
  \BibitemShut {NoStop}%
\bibitem [{\citenamefont {Mart\'inez-Huerta}(2018)}]{Martinez-Huerta:2017gna}%
  \BibitemOpen
  \bibfield  {author} {\bibinfo {author} {\bibfnamefont {H.}~\bibnamefont
  {Mart\'inez-Huerta}} (\bibinfo {collaboration} {HAWC}),\ }\bibfield
  {booktitle} {\emph {\bibinfo {booktitle} {{Contributions to the 35th
  International Cosmic Ray Conference (ICRC 2017)}}},\ }\href {\doibase
  10.22323/1.301.0868} {\bibfield  {journal} {\bibinfo  {journal} {PoS}\
  }\textbf {\bibinfo {volume} {ICRC2017}},\ \bibinfo {pages} {868} (\bibinfo
  {year} {2018})},\ \Eprint {http://arxiv.org/abs/1708.03384} {arXiv:1708.03384
  [astro-ph.HE]} \BibitemShut {NoStop}%
\bibitem [{\citenamefont {Guedes~Lang}\ \emph {et~al.}(2018)\citenamefont
  {Guedes~Lang}, \citenamefont {Mart\'inez-Huerta},\ and\ \citenamefont
  {de~Souza}}]{Lang:2017wpe}%
  \BibitemOpen
  \bibfield  {author} {\bibinfo {author} {\bibfnamefont {R.}~\bibnamefont
  {Guedes~Lang}}, \bibinfo {author} {\bibfnamefont {H.}~\bibnamefont
  {Mart\'inez-Huerta}}, \ and\ \bibinfo {author} {\bibfnamefont
  {V.}~\bibnamefont {de~Souza}},\ }\href {\doibase 10.3847/1538-4357/aa9f2c}
  {\bibfield  {journal} {\bibinfo  {journal} {Astrophys. J.}\ }\textbf
  {\bibinfo {volume} {853}},\ \bibinfo {pages} {23} (\bibinfo {year} {2018})},\
  \Eprint {http://arxiv.org/abs/1701.04865} {arXiv:1701.04865 [astro-ph.HE]}
  \BibitemShut {NoStop}%
\bibitem [{\citenamefont {Cologna}\ \emph {et~al.}(2017)\citenamefont {Cologna}
  \emph {et~al.}}]{Cologna:2016cws}%
  \BibitemOpen
  \bibfield  {author} {\bibinfo {author} {\bibfnamefont {G.}~\bibnamefont
  {Cologna}} \emph {et~al.} (\bibinfo {collaboration} {FACT, H.E.S.S.}),\
  }\bibfield  {booktitle} {\emph {\bibinfo {booktitle} {{Proceedings, 6th
  International Symposium on High-Energy Gamma-Ray Astronomy (Gamma 2016):
  Heidelberg, Germany, July 11-15, 2016}}},\ }\href {\doibase
  10.1063/1.4968965} {\bibfield  {journal} {\bibinfo  {journal} {AIP Conf.
  Proc.}\ }\textbf {\bibinfo {volume} {1792}},\ \bibinfo {pages} {050019}
  (\bibinfo {year} {2017})},\ \Eprint {http://arxiv.org/abs/1611.03983}
  {arXiv:1611.03983 [astro-ph.HE]} \BibitemShut {NoStop}%
\bibitem [{\citenamefont {Lorentz}\ and\ \citenamefont
  {Brun}(2017)}]{Mrk501_HESS_flare}%
  \BibitemOpen
  \bibfield  {author} {\bibinfo {author} {\bibfnamefont {M.}~\bibnamefont
  {Lorentz}}\ and\ \bibinfo {author} {\bibfnamefont {P.}~\bibnamefont {Brun}}
  (\bibinfo {collaboration} {H.E.S.S.}),\ }\bibfield  {booktitle} {\emph
  {\bibinfo {booktitle} {{Proceedings, 6th Roma International Workshop on
  Astroparticle Physics (RICAP16): Rome, Italy, June 21-24, 2016}}},\ }\href
  {\doibase 10.1051/epjconf/201713603018} {\bibfield  {journal} {\bibinfo
  {journal} {EPJ Web Conf.}\ }\textbf {\bibinfo {volume} {136}},\ \bibinfo
  {pages} {03018} (\bibinfo {year} {2017})},\ \Eprint
  {http://arxiv.org/abs/1606.08600} {arXiv:1606.08600 [astro-ph.HE]}
  \BibitemShut {NoStop}%
\bibitem [{\citenamefont {Pfeifer}(2018)}]{Pfeifer:2018pty}%
  \BibitemOpen
  \bibfield  {author} {\bibinfo {author} {\bibfnamefont {C.}~\bibnamefont
  {Pfeifer}},\ }\href {\doibase 10.1016/j.physletb.2018.03.017} {\bibfield
  {journal} {\bibinfo  {journal} {Phys. Lett.}\ }\textbf {\bibinfo {volume}
  {B780}},\ \bibinfo {pages} {246} (\bibinfo {year} {2018})},\ \Eprint
  {http://arxiv.org/abs/1802.00058} {arXiv:1802.00058 [gr-qc]} \BibitemShut
  {NoStop}%
\bibitem [{\citenamefont {Ellis}\ \emph {et~al.}(2018)\citenamefont {Ellis},
  \citenamefont {Konoplich}, \citenamefont {Mavromatos}, \citenamefont
  {Nguyen}, \citenamefont {Sakharov},\ and\ \citenamefont
  {Sarkisyan-Grinbaum}}]{Ellis:2018lca}%
  \BibitemOpen
  \bibfield  {author} {\bibinfo {author} {\bibfnamefont {J.}~\bibnamefont
  {Ellis}}, \bibinfo {author} {\bibfnamefont {R.}~\bibnamefont {Konoplich}},
  \bibinfo {author} {\bibfnamefont {N.~E.}\ \bibnamefont {Mavromatos}},
  \bibinfo {author} {\bibfnamefont {L.}~\bibnamefont {Nguyen}}, \bibinfo
  {author} {\bibfnamefont {A.~S.}\ \bibnamefont {Sakharov}}, \ and\ \bibinfo
  {author} {\bibfnamefont {E.~K.}\ \bibnamefont {Sarkisyan-Grinbaum}},\
  }\href@noop {} {\  (\bibinfo {year} {2018})},\ \Eprint
  {http://arxiv.org/abs/1807.00189} {arXiv:1807.00189 [astro-ph.HE]}
  \BibitemShut {NoStop}%
\bibitem [{\citenamefont {Abdalla}\ and\ \citenamefont
  {Böttcher}(2018)}]{Abdalla:2018sxi}%
  \BibitemOpen
  \bibfield  {author} {\bibinfo {author} {\bibfnamefont {H.}~\bibnamefont
  {Abdalla}}\ and\ \bibinfo {author} {\bibfnamefont {M.}~\bibnamefont
  {Böttcher}},\ }\href {\doibase 10.3847/1538-4357/aadb87} {\bibfield
  {journal} {\bibinfo  {journal} {Astrophys. J.}\ }\textbf {\bibinfo {volume}
  {865}},\ \bibinfo {pages} {159} (\bibinfo {year} {2018})},\ \Eprint
  {http://arxiv.org/abs/1809.00477} {arXiv:1809.00477 [astro-ph.HE]}
  \BibitemShut {NoStop}%
\bibitem [{\citenamefont {Vasileiou}\ \emph {et~al.}(2013)\citenamefont
  {Vasileiou}, \citenamefont {Jacholkowska}, \citenamefont {Piron},
  \citenamefont {Bolmont}, \citenamefont {Couturier}, \citenamefont {Granot},
  \citenamefont {Stecker}, \citenamefont {Cohen-Tanugi},\ and\ \citenamefont
  {Longo}}]{Vasileiou:2013vra}%
  \BibitemOpen
  \bibfield  {author} {\bibinfo {author} {\bibfnamefont {V.}~\bibnamefont
  {Vasileiou}}, \bibinfo {author} {\bibfnamefont {A.}~\bibnamefont
  {Jacholkowska}}, \bibinfo {author} {\bibfnamefont {F.}~\bibnamefont {Piron}},
  \bibinfo {author} {\bibfnamefont {J.}~\bibnamefont {Bolmont}}, \bibinfo
  {author} {\bibfnamefont {C.}~\bibnamefont {Couturier}}, \bibinfo {author}
  {\bibfnamefont {J.}~\bibnamefont {Granot}}, \bibinfo {author} {\bibfnamefont
  {F.~W.}\ \bibnamefont {Stecker}}, \bibinfo {author} {\bibfnamefont
  {J.}~\bibnamefont {Cohen-Tanugi}}, \ and\ \bibinfo {author} {\bibfnamefont
  {F.}~\bibnamefont {Longo}},\ }\href {\doibase 10.1103/PhysRevD.87.122001}
  {\bibfield  {journal} {\bibinfo  {journal} {Phys. Rev.}\ }\textbf {\bibinfo
  {volume} {D87}},\ \bibinfo {pages} {122001} (\bibinfo {year} {2013})},\
  \Eprint {http://arxiv.org/abs/1305.3463} {arXiv:1305.3463 [astro-ph.HE]}
  \BibitemShut {NoStop}%
\bibitem [{\citenamefont {De~Angelis}\ \emph {et~al.}(2013)\citenamefont
  {De~Angelis}, \citenamefont {Galanti},\ and\ \citenamefont
  {Roncadelli}}]{DeAngelis:2013jna}%
  \BibitemOpen
  \bibfield  {author} {\bibinfo {author} {\bibfnamefont {A.}~\bibnamefont
  {De~Angelis}}, \bibinfo {author} {\bibfnamefont {G.}~\bibnamefont {Galanti}},
  \ and\ \bibinfo {author} {\bibfnamefont {M.}~\bibnamefont {Roncadelli}},\
  }\href {\doibase 10.1093/mnras/stt684} {\bibfield  {journal} {\bibinfo
  {journal} {Mon. Not. Roy. Astron. Soc.}\ }\textbf {\bibinfo {volume} {432}},\
  \bibinfo {pages} {3245} (\bibinfo {year} {2013})},\ \Eprint
  {http://arxiv.org/abs/1302.6460} {arXiv:1302.6460 [astro-ph.HE]} \BibitemShut
  {NoStop}%
\bibitem [{\citenamefont {Franceschini}\ \emph {et~al.}(2008)\citenamefont
  {Franceschini}, \citenamefont {Rodighiero},\ and\ \citenamefont
  {Vaccari}}]{Franceschini}%
  \BibitemOpen
  \bibfield  {author} {\bibinfo {author} {\bibfnamefont {A.}~\bibnamefont
  {Franceschini}}, \bibinfo {author} {\bibfnamefont {G.}~\bibnamefont
  {Rodighiero}}, \ and\ \bibinfo {author} {\bibfnamefont {M.}~\bibnamefont
  {Vaccari}},\ }\href {\doibase 10.1051/0004-6361:200809691} {\bibfield
  {journal} {\bibinfo  {journal} {Astron. Astrophys.}\ }\textbf {\bibinfo
  {volume} {487}},\ \bibinfo {pages} {837} (\bibinfo {year} {2008})},\ \Eprint
  {http://arxiv.org/abs/0805.1841} {arXiv:0805.1841 [astro-ph]} \BibitemShut
  {NoStop}%
\bibitem [{\citenamefont {Meyer}(2018)}]{ebltable}%
  \BibitemOpen
  \bibfield  {author} {\bibinfo {author} {\bibfnamefont {M.}~\bibnamefont
  {Meyer}},\ }\href@noop {} {\enquote {\bibinfo {title} {{EBLtable} module},}\
  }\bibinfo {howpublished} {\url{https://github.com/me-manu/ebltable}}
  (\bibinfo {year} {2018}),\ \bibinfo {note} {accessed: 2018-07-20}\BibitemShut
  {NoStop}%
\bibitem [{\citenamefont {Gilmore}\ \emph {et~al.}(2012)\citenamefont
  {Gilmore}, \citenamefont {Somerville}, \citenamefont {Primack},\ and\
  \citenamefont {Dominguez}}]{Gilmore}%
  \BibitemOpen
  \bibfield  {author} {\bibinfo {author} {\bibfnamefont {R.~C.}\ \bibnamefont
  {Gilmore}}, \bibinfo {author} {\bibfnamefont {R.~S.}\ \bibnamefont
  {Somerville}}, \bibinfo {author} {\bibfnamefont {J.~R.}\ \bibnamefont
  {Primack}}, \ and\ \bibinfo {author} {\bibfnamefont {A.}~\bibnamefont
  {Dominguez}},\ }\href {\doibase 10.1111/j.1365-2966.2012.20841.x} {\bibfield
  {journal} {\bibinfo  {journal} {Mon. Not. Roy. Astron. Soc.}\ }\textbf
  {\bibinfo {volume} {422}},\ \bibinfo {pages} {3189} (\bibinfo {year}
  {2012})},\ \Eprint {http://arxiv.org/abs/1104.0671} {arXiv:1104.0671
  [astro-ph.CO]} \BibitemShut {NoStop}%
\bibitem [{\citenamefont {Dominguez}\ \emph {et~al.}(2011)\citenamefont
  {Dominguez} \emph {et~al.}}]{Dominguez}%
  \BibitemOpen
  \bibfield  {author} {\bibinfo {author} {\bibfnamefont {A.}~\bibnamefont
  {Dominguez}} \emph {et~al.},\ }\href {\doibase
  10.1111/j.1365-2966.2010.17631.x} {\bibfield  {journal} {\bibinfo  {journal}
  {Mon. Not. Roy. Astron. Soc.}\ }\textbf {\bibinfo {volume} {410}},\ \bibinfo
  {pages} {2556} (\bibinfo {year} {2011})},\ \Eprint
  {http://arxiv.org/abs/1007.1459} {arXiv:1007.1459 [astro-ph.CO]} \BibitemShut
  {NoStop}%
\bibitem [{\citenamefont {Wakely}\ and\ \citenamefont {Horan}(2018)}]{tevcat}%
  \BibitemOpen
  \bibfield  {author} {\bibinfo {author} {\bibfnamefont {S.}~\bibnamefont
  {Wakely}}\ and\ \bibinfo {author} {\bibfnamefont {D.}~\bibnamefont {Horan}},\
  }\href@noop {} {\enquote {\bibinfo {title} {{TeVCat} catalog},}\ }\bibinfo
  {howpublished} {\url{http://tevcat.uchicago.edu/}} (\bibinfo {year} {2018}),\
  \bibinfo {note} {accessed: 2018-05-31}\BibitemShut {NoStop}%
\bibitem [{\citenamefont {Aharonian}\ \emph {et~al.}(2002)\citenamefont
  {Aharonian} \emph {et~al.}}]{Mrk421_HEGRA_1999}%
  \BibitemOpen
  \bibfield  {author} {\bibinfo {author} {\bibfnamefont {F.}~\bibnamefont
  {Aharonian}} \emph {et~al.} (\bibinfo {collaboration} {HEGRA}),\ }\href
  {\doibase 10.1051/0004-6361:20021005} {\bibfield  {journal} {\bibinfo
  {journal} {Astron. Astrophys.}\ }\textbf {\bibinfo {volume} {393}},\ \bibinfo
  {pages} {89} (\bibinfo {year} {2002})},\ \Eprint
  {http://arxiv.org/abs/astro-ph/0205499} {arXiv:astro-ph/0205499 [astro-ph]}
  \BibitemShut {NoStop}%
\bibitem [{\citenamefont {Aharonian}\ \emph {et~al.}(2005)\citenamefont
  {Aharonian} \emph {et~al.}}]{Mrk421_HESS_2004}%
  \BibitemOpen
  \bibfield  {author} {\bibinfo {author} {\bibfnamefont {F.}~\bibnamefont
  {Aharonian}} \emph {et~al.} (\bibinfo {collaboration} {H.E.S.S.}),\ }\href
  {\doibase 10.1051/0004-6361:20053050} {\bibfield  {journal} {\bibinfo
  {journal} {Astron. Astrophys.}\ }\textbf {\bibinfo {volume} {437}},\ \bibinfo
  {pages} {95} (\bibinfo {year} {2005})},\ \Eprint
  {http://arxiv.org/abs/astro-ph/0506319} {arXiv:astro-ph/0506319 [astro-ph]}
  \BibitemShut {NoStop}%
\bibitem [{\citenamefont {Acciari}\ \emph
  {et~al.}(2011{\natexlab{a}})\citenamefont {Acciari} \emph
  {et~al.}}]{Mrk421_VERITAS_low}%
  \BibitemOpen
  \bibfield  {author} {\bibinfo {author} {\bibfnamefont {V.~A.}\ \bibnamefont
  {Acciari}} \emph {et~al.},\ }\href {\doibase 10.1088/0004-637X/738/1/25}
  {\bibfield  {journal} {\bibinfo  {journal} {Astrophys. J.}\ }\textbf
  {\bibinfo {volume} {738}},\ \bibinfo {pages} {25} (\bibinfo {year}
  {2011}{\natexlab{a}})},\ \Eprint {http://arxiv.org/abs/1106.1210}
  {arXiv:1106.1210 [astro-ph.HE]} \BibitemShut {NoStop}%
\bibitem [{\citenamefont {Sharma}\ \emph {et~al.}(2015)\citenamefont {Sharma},
  \citenamefont {Nayak}, \citenamefont {Koul}, \citenamefont {Bose},
  \citenamefont {Mitra}, \citenamefont {Dhar}, \citenamefont {Tickoo},\ and\
  \citenamefont {Koul}}]{Mrk421_TACTIC_2005}%
  \BibitemOpen
  \bibfield  {author} {\bibinfo {author} {\bibfnamefont {M.}~\bibnamefont
  {Sharma}}, \bibinfo {author} {\bibfnamefont {J.}~\bibnamefont {Nayak}},
  \bibinfo {author} {\bibfnamefont {M.~K.}\ \bibnamefont {Koul}}, \bibinfo
  {author} {\bibfnamefont {S.}~\bibnamefont {Bose}}, \bibinfo {author}
  {\bibfnamefont {A.}~\bibnamefont {Mitra}}, \bibinfo {author} {\bibfnamefont
  {V.~K.}\ \bibnamefont {Dhar}}, \bibinfo {author} {\bibfnamefont {A.~K.}\
  \bibnamefont {Tickoo}}, \ and\ \bibinfo {author} {\bibfnamefont
  {R.}~\bibnamefont {Koul}},\ }\href {\doibase 10.1016/j.nima.2014.10.012}
  {\bibfield  {journal} {\bibinfo  {journal} {Nucl. Instrum. Meth.}\ }\textbf
  {\bibinfo {volume} {A770}},\ \bibinfo {pages} {42} (\bibinfo {year}
  {2015})},\ \Eprint {http://arxiv.org/abs/1410.2260} {arXiv:1410.2260
  [astro-ph.HE]} \BibitemShut {NoStop}%
\bibitem [{\citenamefont {Chandra}\ \emph {et~al.}(2012)\citenamefont {Chandra}
  \emph {et~al.}}]{Mrk421_TACTIC_2009}%
  \BibitemOpen
  \bibfield  {author} {\bibinfo {author} {\bibfnamefont {P.}~\bibnamefont
  {Chandra}} \emph {et~al.},\ }\href {\doibase 10.1088/0954-3899/39/4/045201}
  {\bibfield  {journal} {\bibinfo  {journal} {J. Phys.}\ }\textbf {\bibinfo
  {volume} {G39}},\ \bibinfo {pages} {045201} (\bibinfo {year} {2012})},\
  \Eprint {http://arxiv.org/abs/1202.2984} {arXiv:1202.2984 [astro-ph.HE]}
  \BibitemShut {NoStop}%
\bibitem [{\citenamefont {Godambe}\ \emph {et~al.}(2008)\citenamefont {Godambe}
  \emph {et~al.}}]{Mrk501_TACTIC}%
  \BibitemOpen
  \bibfield  {author} {\bibinfo {author} {\bibfnamefont {S.~V.}\ \bibnamefont
  {Godambe}} \emph {et~al.},\ }\href {\doibase 10.1088/0954-3899/35/6/065202}
  {\bibfield  {journal} {\bibinfo  {journal} {J. Phys.}\ }\textbf {\bibinfo
  {volume} {G35}},\ \bibinfo {pages} {065202} (\bibinfo {year} {2008})},\
  \Eprint {http://arxiv.org/abs/0804.1473} {arXiv:0804.1473 [astro-ph]}
  \BibitemShut {NoStop}%
\bibitem [{\citenamefont {Bartoli}\ \emph {et~al.}(2012)\citenamefont {Bartoli}
  \emph {et~al.}}]{Mrk501_ARGO_flare}%
  \BibitemOpen
  \bibfield  {author} {\bibinfo {author} {\bibfnamefont {B.}~\bibnamefont
  {Bartoli}} \emph {et~al.} (\bibinfo {collaboration} {ARGO-YBJ}),\ }\href
  {\doibase 10.1088/0004-637X/758/1/2} {\bibfield  {journal} {\bibinfo
  {journal} {Astrophys. J.}\ }\textbf {\bibinfo {volume} {758}},\ \bibinfo
  {pages} {2} (\bibinfo {year} {2012})},\ \Eprint
  {http://arxiv.org/abs/1209.0534} {arXiv:1209.0534 [astro-ph.HE]} \BibitemShut
  {NoStop}%
\bibitem [{\citenamefont {Daniel}\ \emph {et~al.}(2005)\citenamefont {Daniel}
  \emph {et~al.}}]{s1ES1959_Whipple}%
  \BibitemOpen
  \bibfield  {author} {\bibinfo {author} {\bibfnamefont {M.~K.}\ \bibnamefont
  {Daniel}} \emph {et~al.} (\bibinfo {collaboration} {VERITAS}),\ }\href
  {\doibase 10.1086/427406} {\bibfield  {journal} {\bibinfo  {journal}
  {Astrophys. J.}\ }\textbf {\bibinfo {volume} {621}},\ \bibinfo {pages} {181}
  (\bibinfo {year} {2005})},\ \Eprint {http://arxiv.org/abs/astro-ph/0503085}
  {arXiv:astro-ph/0503085 [astro-ph]} \BibitemShut {NoStop}%
\bibitem [{\citenamefont {Aharonian}\ \emph
  {et~al.}(2003{\natexlab{a}})\citenamefont {Aharonian}, \citenamefont
  {Akhperjanian},\ and\ \citenamefont {Beilicke}}]{s1ES1959_HEGRA_low}%
  \BibitemOpen
  \bibfield  {author} {\bibinfo {author} {\bibfnamefont {F.}~\bibnamefont
  {Aharonian}}, \bibinfo {author} {\bibfnamefont {A.}~\bibnamefont
  {Akhperjanian}}, \ and\ \bibinfo {author} {\bibfnamefont {M.}~\bibnamefont
  {Beilicke}} (\bibinfo {collaboration} {HEGRA}),\ }\href {\doibase
  10.1051/0004-6361:20030838} {\bibfield  {journal} {\bibinfo  {journal}
  {Astron. Astrophys.}\ }\textbf {\bibinfo {volume} {406}},\ \bibinfo {pages}
  {L9} (\bibinfo {year} {2003}{\natexlab{a}})},\ \Eprint
  {http://arxiv.org/abs/astro-ph/0305275} {arXiv:astro-ph/0305275 [astro-ph]}
  \BibitemShut {NoStop}%
\bibitem [{\citenamefont {Aharonian}\ \emph
  {et~al.}(2003{\natexlab{b}})\citenamefont {Aharonian} \emph
  {et~al.}}]{H1426_HEGRA_1999}%
  \BibitemOpen
  \bibfield  {author} {\bibinfo {author} {\bibfnamefont {F.}~\bibnamefont
  {Aharonian}} \emph {et~al.} (\bibinfo {collaboration} {HEGRA}),\ }\href
  {\doibase 10.1051/0004-6361:20030326} {\bibfield  {journal} {\bibinfo
  {journal} {Astron. Astrophys.}\ }\textbf {\bibinfo {volume} {403}},\ \bibinfo
  {pages} {523} (\bibinfo {year} {2003}{\natexlab{b}})},\ \Eprint
  {http://arxiv.org/abs/astro-ph/0301437} {arXiv:astro-ph/0301437 [astro-ph]}
  \BibitemShut {NoStop}%
\bibitem [{\citenamefont {Aharonian}\ \emph
  {et~al.}(2007{\natexlab{a}})\citenamefont {Aharonian} \emph
  {et~al.}}]{s1ES0229_HESS}%
  \BibitemOpen
  \bibfield  {author} {\bibinfo {author} {\bibfnamefont {F.}~\bibnamefont
  {Aharonian}} \emph {et~al.} (\bibinfo {collaboration} {H.E.S.S.}),\ }\href
  {\doibase 10.1051/0004-6361:20078462} {\bibfield  {journal} {\bibinfo
  {journal} {Astron. Astrophys.}\ }\textbf {\bibinfo {volume} {475}},\ \bibinfo
  {pages} {L9} (\bibinfo {year} {2007}{\natexlab{a}})},\ \Eprint
  {http://arxiv.org/abs/0709.4584} {arXiv:0709.4584 [astro-ph]} \BibitemShut
  {NoStop}%
\bibitem [{\citenamefont {Aliu}\ \emph {et~al.}(2014)\citenamefont {Aliu} \emph
  {et~al.}}]{s1ES0229_VERITAS}%
  \BibitemOpen
  \bibfield  {author} {\bibinfo {author} {\bibfnamefont {E.}~\bibnamefont
  {Aliu}} \emph {et~al.},\ }\href {\doibase 10.1088/0004-637X/782/1/13}
  {\bibfield  {journal} {\bibinfo  {journal} {Astrophys. J.}\ }\textbf
  {\bibinfo {volume} {782}},\ \bibinfo {pages} {13} (\bibinfo {year} {2014})},\
  \Eprint {http://arxiv.org/abs/1312.6592} {arXiv:1312.6592 [astro-ph.HE]}
  \BibitemShut {NoStop}%
\bibitem [{\citenamefont {Aharonian}\ \emph
  {et~al.}(2007{\natexlab{b}})\citenamefont {Aharonian} \emph
  {et~al.}}]{s1ES0347_VERITAS}%
  \BibitemOpen
  \bibfield  {author} {\bibinfo {author} {\bibfnamefont {F.}~\bibnamefont
  {Aharonian}} \emph {et~al.} (\bibinfo {collaboration} {H.E.S.S.}),\ }\href
  {\doibase 10.1051/0004-6361:20078412} {\bibfield  {journal} {\bibinfo
  {journal} {Astron. Astrophys.}\ }\textbf {\bibinfo {volume} {473}},\ \bibinfo
  {pages} {L25} (\bibinfo {year} {2007}{\natexlab{b}})},\ \Eprint
  {http://arxiv.org/abs/0708.3021} {arXiv:0708.3021 [astro-ph]} \BibitemShut
  {NoStop}%
\bibitem [{\citenamefont {Bartoli}\ \emph {et~al.}(2011)\citenamefont {Bartoli}
  \emph {et~al.}}]{Mrk421_ARGO_flux1}%
  \BibitemOpen
  \bibfield  {author} {\bibinfo {author} {\bibfnamefont {B.}~\bibnamefont
  {Bartoli}} \emph {et~al.} (\bibinfo {collaboration} {ARGO-YBJ}),\ }\href
  {\doibase 10.1088/0004-637X/734/2/110} {\bibfield  {journal} {\bibinfo
  {journal} {Astrophys. J.}\ }\textbf {\bibinfo {volume} {734}},\ \bibinfo
  {pages} {110} (\bibinfo {year} {2011})},\ \Eprint
  {http://arxiv.org/abs/1106.0896} {arXiv:1106.0896 [astro-ph.HE]} \BibitemShut
  {NoStop}%
\bibitem [{\citenamefont {Schroedter}\ \emph {et~al.}(2005)\citenamefont
  {Schroedter} \emph {et~al.}}]{s1ES2344_Whipple_2005_b}%
  \BibitemOpen
  \bibfield  {author} {\bibinfo {author} {\bibfnamefont {M.}~\bibnamefont
  {Schroedter}} \emph {et~al.},\ }\href {\doibase 10.1086/496968} {\bibfield
  {journal} {\bibinfo  {journal} {Astrophys. J.}\ }\textbf {\bibinfo {volume}
  {634}},\ \bibinfo {pages} {947} (\bibinfo {year} {2005})},\ \Eprint
  {http://arxiv.org/abs/astro-ph/0508499} {arXiv:astro-ph/0508499 [astro-ph]}
  \BibitemShut {NoStop}%
\bibitem [{\citenamefont {Acciari}\ \emph
  {et~al.}(2011{\natexlab{b}})\citenamefont {Acciari} \emph
  {et~al.}}]{s1ES2344_VERITAS_low}%
  \BibitemOpen
  \bibfield  {author} {\bibinfo {author} {\bibfnamefont {V.~A.}\ \bibnamefont
  {Acciari}} \emph {et~al.},\ }\href {\doibase 10.1088/0004-637X/738/2/169}
  {\bibfield  {journal} {\bibinfo  {journal} {Astrophys. J.}\ }\textbf
  {\bibinfo {volume} {738}},\ \bibinfo {pages} {169} (\bibinfo {year}
  {2011}{\natexlab{b}})},\ \Eprint {http://arxiv.org/abs/1106.4594}
  {arXiv:1106.4594 [astro-ph.HE]} \BibitemShut {NoStop}%
\bibitem [{\citenamefont {Allen}\ \emph {et~al.}(2017)\citenamefont {Allen}
  \emph {et~al.}}]{s1ES2344_VERITAS_2017}%
  \BibitemOpen
  \bibfield  {author} {\bibinfo {author} {\bibfnamefont {C.}~\bibnamefont
  {Allen}} \emph {et~al.} (\bibinfo {collaboration} {VERITAS}),\ }\href
  {\doibase 10.1093/mnras/stx1756} {\bibfield  {journal} {\bibinfo  {journal}
  {Mon. Not. Roy. Astron. Soc.}\ }\textbf {\bibinfo {volume} {471}},\ \bibinfo
  {pages} {2117} (\bibinfo {year} {2017})},\ \Eprint
  {http://arxiv.org/abs/1708.02829} {arXiv:1708.02829 [astro-ph.HE]}
  \BibitemShut {NoStop}%
\bibitem [{\citenamefont {Aliu}\ \emph {et~al.}(2013)\citenamefont {Aliu} \emph
  {et~al.}}]{s1ES1959_VERITAS}%
  \BibitemOpen
  \bibfield  {author} {\bibinfo {author} {\bibfnamefont {E.}~\bibnamefont
  {Aliu}} \emph {et~al.} (\bibinfo {collaboration} {VERITAS}),\ }\href
  {\doibase 10.1088/0004-637X/775/1/3} {\bibfield  {journal} {\bibinfo
  {journal} {Astrophys. J.}\ }\textbf {\bibinfo {volume} {775}},\ \bibinfo
  {pages} {3} (\bibinfo {year} {2013})},\ \Eprint
  {http://arxiv.org/abs/1307.6772} {arXiv:1307.6772 [astro-ph.HE]} \BibitemShut
  {NoStop}%
\bibitem [{\citenamefont {Santander}(2017)}]{s1ES1959_VERITAS_2015}%
  \BibitemOpen
  \bibfield  {author} {\bibinfo {author} {\bibfnamefont {M.}~\bibnamefont
  {Santander}} (\bibinfo {collaboration} {VERITAS}),\ }\bibfield  {booktitle}
  {\emph {\bibinfo {booktitle} {{Proceedings, 35th International Cosmic Ray
  Conference (ICRC 2017): Bexco, Busan, Korea, July 12-20, 2017}}},\
  }\href@noop {} {\bibfield  {journal} {\bibinfo  {journal} {PoS}\ }\textbf
  {\bibinfo {volume} {ICRC2017}},\ \bibinfo {pages} {622} (\bibinfo {year}
  {2017})},\ \Eprint {http://arxiv.org/abs/1709.02365} {arXiv:1709.02365
  [astro-ph.HE]} \BibitemShut {NoStop}%
\bibitem [{\citenamefont {Martínez-Huerta}\ \emph {et~al.}(2019)\citenamefont
  {Martínez-Huerta}, \citenamefont {Lang},\ and\ \citenamefont
  {de~Souza}}]{Martinez-Huerta:2019ehp}%
  \BibitemOpen
  \bibfield  {author} {\bibinfo {author} {\bibfnamefont {H.}~\bibnamefont
  {Martínez-Huerta}}, \bibinfo {author} {\bibfnamefont {R.~G.}\ \bibnamefont
  {Lang}}, \ and\ \bibinfo {author} {\bibfnamefont {V.}~\bibnamefont
  {de~Souza}},\ }in\ \href@noop {} {\emph {\bibinfo {booktitle} {{International
  Conference on Black Holes as Cosmic Batteries: UHECRs and Multimessenger
  Astronomy (BHCB 2018) Foz do Iguaçu, Brazil, September 12-15, 2018}}}}\
  (\bibinfo {year} {2019})\ \Eprint {http://arxiv.org/abs/1901.03205}
  {arXiv:1901.03205 [astro-ph.HE]} \BibitemShut {NoStop}%
\end{thebibliography}%

\end{document}